\documentclass[sigconf]{acmart}

\usepackage{graphicx}    % Required for graphics (e.g., in table cells)
\usepackage{multirow}    % Enables multi-row cells in tables
\usepackage{makecell}    % Allows line breaks and formatting in table cells
\usepackage{array}       % Improves column formatting in tables
\usepackage{booktabs}    % Enhances table appearance with better lines
\usepackage{caption}     % Adjusts table captions
\usepackage{float}       % Provides control over table placement
\usepackage{xspace}
\usepackage{enumitem}
\usepackage{color}
\usepackage{pifont}

\newcommand{\sys}{\textit{\sc MIKU}\xspace}
\AtBeginDocument{%
  }

\setcopyright{acmlicensed}
\copyrightyear{2018}
\acmYear{2018}
\acmDOI{XXXXXXX.XXXXXXX}

\begin{document}
\title{Architectural and System Implications of CXL-enabled \\
Tiered Memory}

% \title{Contention in CXL-DRAM Co-running System}

% \subtitle{ISCA 2025 Submission $\#996$ -- Confidential Draft -- Do NOT Distribute}

\author{Yujie Yang}
\affiliation{%
  \institution{The University of Texas at Arlington}
  \country{}
}
\email{yujie.yang@uta.edu}

\author{Lingfeng Xiang}
\affiliation{%
  \institution{The University of Texas at Arlington}
  \country{}
}
\email{lingfeng.xiang@mavs.uta.edu}

\author{Peiran Du}
\affiliation{%
  \institution{The University of Texas at Arlington}
    \country{}
}
\email{dupeiran3@gmail.com}

\author{Zhen Lin}
\affiliation{%
  \institution{The University of Texas at Arlington}
    \country{}
}
\email{zxl5722@mavs.uta.edu}

\author{Weishu Deng}
\affiliation{%
  \institution{The University of Texas at Arlington}
    \country{}
}
\email{weishu.deng@mavs.uta.edu}

\author{Ren Wang}
\affiliation{%
  \institution{Intel Labs}
    \country{}
}
\email{ren.wang@intel.com}

\author{Andrey Kudryavtsev}
\affiliation{%
  \institution{Micron}
    \country{}
}
\email{akudryavtsev@micron.com}

\author{Louis Ko}
\affiliation{%
  \institution{Supermicro}
    \country{}
}
\email{louisk@supermicro.com}

\author{Hui Lu}
\affiliation{%
  \institution{The University of Texas at Arlington}
    \country{}
}
\email{hui.lu@uta.edu}

\author{Jia Rao}
\affiliation{%
  \institution{The University of Texas at Arlington}
    \country{}
}
\email{jia.rao@uta.edu}

%\author{ISCA2025 Submission #NaN -- ConfidentialDraft -- Do NOT Distribute‼}


\begin{abstract}
Memory disaggregation is an emerging technology that decouples memory from traditional memory buses, enabling independent scaling of compute and memory. Compute Express Link (CXL), an open-standard interconnect technology, facilitates memory disaggregation by allowing processors to access remote memory through the PCIe bus while preserving the shared-memory programming model. This innovation creates a tiered memory architecture combining local DDR and remote CXL memory with distinct performance characteristics.

In this paper, we investigate the architectural implications of CXL memory, focusing on its increased latency and performance heterogeneity, which can undermine the efficiency of existing processor designs optimized for (relatively) uniform memory latency. Using carefully designed micro-benchmarks, we identify bottlenecks such as limited hardware-level parallelism in CXL memory, unfair queuing in memory request handling, and its impact on DDR memory performance and inter-core synchronization. Our findings reveal that the disparity in memory tier parallelism can reduce DDR memory bandwidth by up to 81\% under heavy loads. To address these challenges, we propose a Dynamic Memory Request Control mechanism, \sys, that prioritizes DDR memory requests while serving CXL memory requests on a best-effort basis. By dynamically adjusting CXL request rates based on service time estimates, \sys achieves near-peak DDR throughput while maintaining high performance for CXL memory. Our evaluation with micro-benchmarks and representative workloads demonstrates the potential of \sys to enhance tiered memory system efficiency.
\end{abstract}

\maketitle

\section{Introduction}
Memory disaggregation is an emerging technology that decouples memory from the traditional memory bus, allowing computing and memory to scale independently. This new architecture will enable processors across hosts to share a common memory pool, resulting in higher memory utilization, improved cost efficiency, and greatly expanded memory capacity for memory-intensive applications. Compute Express Link (CXL)~\cite{cxl, memtis, Maruf_23asplos_tpp} is an open-standard interconnect technology, jointly developed by Intel, AMD, NVIDIA, Micron, and many other key players in the tech industry, to support memory disaggregation. CXL enables processors to access memory devices through the PCIe bus, either within a single host or across multiple hosts interconnected by a PCIe switch. Most importantly, CXL preserves the conventional shared-memory programming interface that allows unmodified applications to access disaggregated memory in a byte-addressable manner using ordinary {\tt load}/{\tt store} instructions. CXL is a significant architectural innovation that seeks to unify (CPU-attached) host memory, device memory (e.g., GPU memory and data processing units (DPU) memory), and any byte-addressable devices on the PCIe bus (e.g., DRAM, persistent memory, and SSDs) in a cache-coherent domain. 

As CXL technology continues to advance, commercial CXL devices are emerging in various forms, including FPGA-based memory expanders~\cite{fpga-intel}, ASIC-based memory modules for single-host memory expansion~\cite{microcxl,samsungcxl}, and CXL-enabled servers for multi-host memory pooling~\cite{h3cxl}. Together with traditional DRAM (local) memory directly attached to the processor, CXL-enabled disaggregated (remote) memory enables a tiered memory architecture comprising memory devices with distinct characteristics, including speed, size, power, and cost. Since the commercial availability of CXL memory, studies have emerged characterizing its behavior and examining its impact on application performance~\cite{cxl_study, liu2024dissectingcxlmemoryperformance, cxl-perf-models, liu2024exploringevaluatingrealworldcxl}. Most studies have recognized the overhead introduced by the CXL protocol and the increased data access latency over the PCIe bus. However, they often treated CXL memory as an isolated, black-box device, relying on extensive empirical analyses and statistical modeling to understand performance slowdowns. A recent study~\cite{liu2024dissectingcxlmemoryperformance} has reported that CXL memory's high latency can reduce the effectiveness of CPU prefetchers. However, many architectural implications of tiered memory systems remain unexplored.

In this paper, we seek to validate a hypothesis -- {\em The increased memory latency and performance heterogeneity introduced by CXL memory can undermine the effectiveness and efficiency of existing processor designs, which are built on the assumption of low and relatively uniform memory latency in traditional memory hierarchies}. Unlike existing studies that use real applications to evaluate the impact of CXL memory on various workloads, we employ carefully designed micro-benchmarks to identify potential architectural bottlenecks. Specifically, we employ a latency benchmark ({\tt lat-test}) designed to minimize instruction and memory-level parallelism to precisely quantify the overhead introduced by the CXL protocol and PCIe bus. Additionally, we use a bandwidth benchmark ({\tt bw-test}) to stress-test CXL memory by fully utilizing the processor's concurrent memory request handling capabilities and the hardware parallelism of the CXL device. We also explore various memory configurations, including the ratio between traditional DDR and CXL memory, as well as hardware and software memory interleaving, on two platforms from different vendors to identify potential architectural bottlenecks.

Through rigorous benchmarking and profiling, we have made important discoveries that can help guide the design of future tiered memory systems and help identify applications that benefit most from tiered memory. {\em First}, while CXL memory incurs inherently higher latency compared to local DDR memory, it behaves similarly to traditional memory in terms of average and tail latency if its underlying medium is DDR memory. {\em Second}, significant performance discrepancies between CXL memory and local DDR memory can occur due to limited hardware-level parallelism in CXL memory. This limitation restricts per-device CXL memory bandwidth and results in a substantial increase in latency, up to 8-10 times higher. The CXL devices we tested exhibit peak bandwidth and hardware parallelism comparable to a single DDR DIMM, despite offering 4-8x the capacity of a DDR DIMM. {\em Third}, modern processors, by default, interleave (via hardware) data access across all DDR DIMMs within a socket, effectively combining per-DIMM parallelism to serve concurrent memory requests. Thus, the disparity in parallelism between local DDR memory and remote CXL memory becomes more pronounced. The disparity in handling concurrent memory requests, particularly under heavy load, between memory tiers leads to unfair memory request processing for the faster memory tier. This imbalance can significantly reduce DDR memory bandwidth, by as much as 81\%. This issue has been observed in both Vendor A's Caching and Home Agent (CHA) and Vendor B's Core Complex (CCX). {\em Fourth}, the unfair queuing impacts not only memory request processing but also the effectiveness of the last-level cache (LLC) in handling cache hits and the efficiency of inter-core synchronization.

The fundamental issue is that the upstream components in a processor data pipeline, such as cores and prefetchers, issue data requests at a uniform rate to tiered memory devices where these requests are serviced at distinct rates due to performance heterogeneity. Flow control indiscriminately throttles all memory requests when intermediate request queues (e.g., CHA and CCX) become full only after a disproportionate backlog of requests for CXL memory has built up in the queues, permanently affecting the dispatch rate of requests for fast DDR memory. To address this issue, we propose a \underline{D}ynamic \underline{M}emory \underline{R}equest \underline{C}ontrol mechanism, named {\em \sys}, to prioritize DDR memory requests while serving CXL memory requests in a best-effort basis. The goal is to preserve the {\em work-conserving} property in memory serving and aim to combine the bandwidths of both DDR and CXL memory. \sys employs the estimated CXL memory service time, which is measured using hardware performance counters, as a measure for request backlog, and dynamically adjusts the issue rate of CXL memory requests. Evaluation with micro-benchmarks and three representative memory-intensive workloads show that \sys can effectively prioritize local DDR memory requests, closely approaching its peak throughput, while maintaining high performance for CXL memory.

\section{Background and Motivation}

%\begin{figure}[t]
%\centering
  
%  \includegraphics[width=0.45\textwidth]{figures/cxl-arch.pdf}
 
%  \caption{Three types of CXL devices. CXL type 3 devices are mainly used for memory %expansion/scaling.}
%  \label{fig:cxl-arch}
%\end{figure}
Compute Express Link (CXL) is an open-standard, high-performance interconnect technology designed to facilitate efficient communication between processors, memory, and accelerators, such as GPUs, DPUs, and FPGAs, using the PCIe bus. CXL includes three primary protocols to enable byte-addressable memory access and sharing across connected devices. The {\tt CXL.io} protocol handles device discovery, configuration, and management, ensuring compatibility with existing PCIe infrastructure. The {\tt CXL.cache} protocol enables coherent caching between the host processor (CPU) and attached CXL devices. The {\tt CXL.mem} protocol facilitates direct memory access between the host CPU and CXL devices. 

%Figure~\ref{fig:cxl-arch} shows three types of CXL devices and their associated CXL protocols. Of the three types of CXL devices, Type 2 and Type 3 devices include on-device memory that can be directly accessed by the host processor. While the {\tt CXL.cache} protocol has significant potential to reduce data movement between the host processor and accelerators, as of this writing, no commercial CXL Type 2 products fully implement {\tt CXL.cache} due to concerns about coherency management overhead and hardware complexity. In contrast, the {\tt CXL.mem} protocol enables a host to expand memory via compute-less CXL type 3 devices, as shown in Figure~\ref{fig:cxl-arch} (c). Thus, the host processor can access both its locally attached DDR memory and the remote CXL memory, which may include DDR memory, persistent memory, and SSDs, using {\tt load} and {\tt store} instructions. Type 3 CXL memory devices are commercially available, typically in the form of FPGA-based memory expanders, ASIC-based memory modules for single-host systems, and CXL-enabled servers for multi-host memory pooling. 

\begin{figure}[t]
\centering
  
  \includegraphics[width=0.28\textwidth]{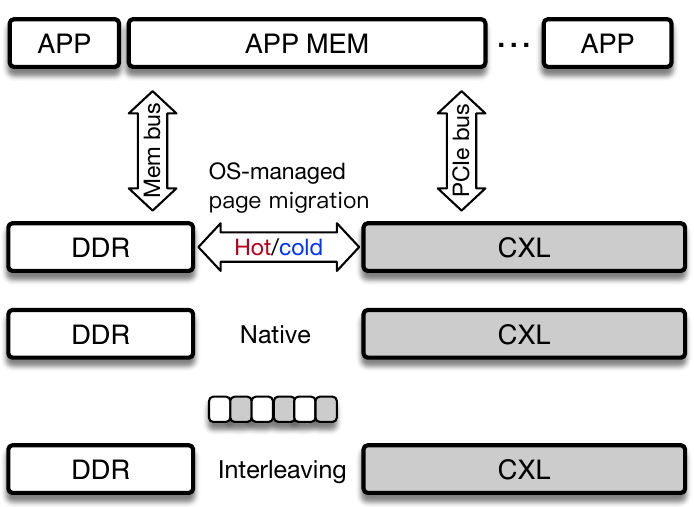}
  \vspace{-0.1in}
  \caption{Application utilizes tiered memory through three memory management schemes.}
  \vspace{-0.2in}
  \label{fig:cxl-mem}
\end{figure}
Since CXL memory retains the traditional byte-addressable programming interface, it can be managed by the operating system (OS) and allocated to applications as regular memory while offering distinct performance, cost, and capacity trade-offs. While there are existing studies~\cite{liu2024dissectingcxlmemoryperformance, cxl_study, liu2024exploringevaluatingrealworldcxl} that characterized the performance of CXL memory compared to traditional DRAM, they mainly focused on quantifying the overhead of the CXL protocol and the efficiency of PCIe-based memory access. There is limited understanding of how CXL-enabled tiered memory, the resulting memory heterogeneity, especially performance heterogeneity, will interact with the processor, the CPU cache, and the remaining memory hierarchy. Additionally, there are ongoing debates in the industry and academia on how applications can effectively utilize tiered memory. 

Figure~\ref{fig:cxl-mem} shows three different ways that applications can use tiered memory. {\em OS-managed} delegates tiered memory management to the OS, which dynamically migrates memory pages between local (DDR) memory and remote (CXL) memory based on page hotness. Users are abstracted from the tiered memory layout and rely on the OS to automatically move hot data to faster memory for improved performance. {\em Native} exposes the underlying tiered memory layout to users, enabling them to implement application-specific optimizations for memory management. {\em Interleaving} distributes consecutive memory addresses across DDR and CXL memory at the page level. Memory interleaving can potentially enhance memory throughput, improve utilization by balancing the workload across different memory types, and reduce fragmentation, thereby improving overall system performance.

\begin{figure}[t]
\centering
  
  \includegraphics[width=0.38\textwidth]{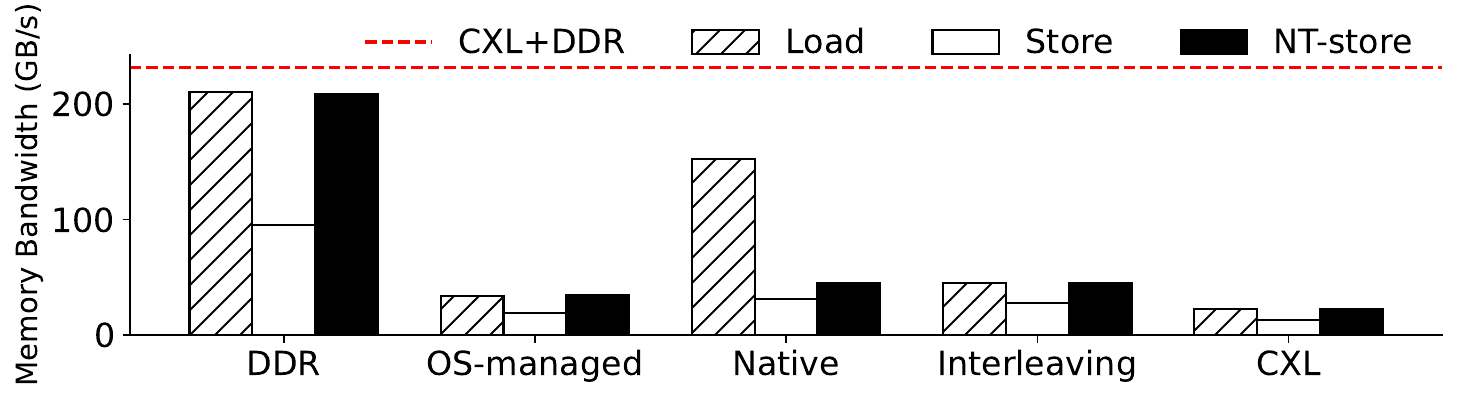}
  \caption{State-of-the-art tiered memory management schemes fall short of achieving the expected combined bandwidth of DDR and CXL.}
  \label{fig:cxl-mem-profiling}
\end{figure}
Unfortunately, it is non-trivial to attain high performance via memory tiering. We used a carefully crafted micro-benchmark to measure the sustained memory bandwidth on three types of tiered memory testbeds (see Section~\S\ref{sec:profiling} for details on hardware configuration). The benchmark's working set size (WSS) can fit into the fast DDR memory, in which case we obtained the maximally achievable (upper bound) bandwidth. We also measured the benchmark performance when its WSS was placed entirely in slow CXL memory, which serves as the baseline (lower bound) performance. For the other tiered memory management schemes, we ran two copies of the benchmark -- one on DDR memory and the other on CXL memory. Figure~\ref{fig:cxl-mem-profiling} shows the aggregated bandwidth of the two benchmarks for the three tiered memory management schemes on a latest platform from vendor A. Ideally, the aggregated bandwidth should closely match the sum of DDR and CXL bandwidths, represented by the red dotted line. However, all three tiered memory management schemes experienced significant performance loss, particularly for {\tt stores} and {\tt non-temporal stores}. Recent studies~\cite{Xiang-OSDI24, Maruf_23asplos_tpp, liu2024dissectingcxlmemoryperformance} have shown that page migrations can significantly degrade tiered memory performance, which accounts for the low bandwidth observed in {\em OS-managed}. Nevertheless, the underlying causes of bandwidth loss in {\em native} and {\em interleaving} remain to be investigated.

In this paper, we aim to develop a comprehensive and in-depth understanding of the architectural interactions between CXL memory, the host processor, and the broader memory hierarchy. Our hypothesis is that the significantly higher latency of CXL memory, together with the performance heterogeneity it introduces, will lead to inefficiencies in processors that are designed and optimized for relatively uniform memory latencies and a single type of DDR memory. 
Recent research~\cite{liu2024dissectingcxlmemoryperformance} has revealed that the high latency of CXL memory reduces the effectiveness of cacheline prefetchers, subsequently degrading CPU cache efficiency. 

\section{Profiling Configurations}
\label{sec:profiling}
\noindent{\bf Platforms.}
We conducted experiments on {\em two} platforms with varying configurations in host CPUs, local DRAM, and remote CXL memory, as detailed in Table~\ref{tab:config}. 

\begin{itemize}[leftmargin=*]
\setlength{\itemindent}{0em}
\setlength\itemsep{0em}
\item {\em Platform A} was equipped with two sockets, each featuring a processor from vendor A, 8 DDR DIMMs (32GB each), and two 256GB Micron's (pre-market) CXL memory devices. 

\item {\em Platform B} had a single socket equipped with a processor from vendor B, 12 DDR DIMMs (16GB each), and four 256GB Micron's (pre-market) CXL memory devices. 
\end{itemize}

%By default, both platforms have all DDR and CXL memory slots fully populated. 
On both platforms, hardware interleaving was applied to all DDRs, while CXL memory devices were interleaved in software using Linux's weighted interleaving~\cite{kernel-interleaving}, with equal weights.%Table~\ref{tab:config} also lists the DDR and CXL memory performance characteristics of the two platforms for single-threaded and peak (multi-threaded) performance, respectively. More discussions are in Section~\S\ref{sec:quant}.

\begin{table}[t]
\centering
\setlength{\belowcaptionskip}{0pt} % Space below the caption
\footnotesize
\begin{tabular}{ c | c | c }
  \hline
  \thead{} & Platform A  & Platform B\\
  \hline
  \hline
  CPUs & \makecell{From Vendor A} & \makecell{From Vendor B} \\
  \hline
  DRAM & \makecell{8 * 32 GB DDR5 (4800MT/s)} & \makecell{12 * 16 GB DDR5 (4800MT/s)} \\
  \hline
  CXL memory & \makecell{2 * 256 GB PCIe Gen5x8} & \makecell{4 * 256 GB PCIe Gen5x8} \\
  \hline
\end{tabular}
\caption{The configurations of two platforms.}
  \vspace{-0.2in}

%and performance characteristics of various memory devices.  }
\label{tab:config}
\end{table}

\noindent{\bf Benchmarks.}
We adopted two types of benchmarks, for memory bandwidth ({\tt bw-test}) and latency ({\tt lat-test}) measurement, to analyze memory heterogeneity.

\noindent{\bf \em Bandwidth tests.}
We evaluated memory bandwidth for read and write operations using lmbench~\cite{larry_96atc_lmbench}. To conduct the tests, we launched varying numbers of threads, each assigned a 1 GB {\em non-overlapping} memory region, performing sequential read and write operations within their respective address spaces. Furthermore, we measured non-temporal write ({\tt nt-store}) bandwidth to estimate the upper bandwidth limit of the hardware. Since lmbench does not include tests for non-temporal writes, we developed our own. Like lmbench, our microbenchmark used manually unrolled loops with AVX non-temporal write instructions.

\noindent{\bf \em Latency tests.}
%\textcolor{red}{Latency test description.} 
Inspired by ~\cite{drepper2007every}, we measured memory latency using pointer chasing. Within a given address range, we placed a pointer on each cacheline. During initialization, we connected these pointers in a randomly shuffled order, forming a circular singly linked list. We ensured a stable average latency by configuring the memory address range significantly larger than the LLC and repeating the access pattern multiple times.

We measured memory access tail latency by comparing per-access latency against a predefined threshold. Specifically, we reserved two registers: one to store the timestamp of the previous memory access and another to count accesses with latency below the threshold. For each access, we retrieved the current time using the Time Stamp Counter (TSC), subtracted the previous timestamp, and compared the resulting duration with the threshold. Depending on the result, we updated the registers -- replacing the previous timestamp with the current time and incrementing the counter if the duration was below the threshold. After completing the traversal, we calculated the percentage of accesses below the threshold to represent the tail latency percentage. By varying the latency threshold, we obtained the memory latency distribution. Given the high latency of memory accesses (100s of CPU cycles), the overhead from timestamp subtraction and counter updates was negligible. 
%\textcolor{red}{Tail latency test description.} 
%We collected memory access tail latency by comparing per-access latency against a statically defined threshold.
%with minimum storage overhead. 
%Specifically, we reserved two registers, one for the previous pointer access timestamp and one for less-than-threshold access counts. We subtracted the previous timestamp register from the current time retrieved from Time Stamp Counter~(TSC), compared the duration with the threshold, and updated the two registers accordingly -- i.e., replacing the timestamp register with the current time and incrementing the access counter register if the duration is below the threshold. After the traversal, we calculated the percentage of accesses below the threshold, representing the tail latency percentage. We obtained the memory latency distribution by varying the defined latency threshold. Given the high latency of memory accesses (hundreds of CPU cycles), the overhead of timestamp subtraction and counter updates was negligible.

\noindent{\bf Performance monitoring.}
We leveraged Intel Performance Monitoring Units (PMUs\cite{intelpmu}) to capture architectural events, focusing specifically on events related to memory requests, bandwidth, and other potential bottlenecks in the system. To facilitate this process, we employed existing tools such as Intel Performance Counter Monitor (PCM)~\cite{pcm, intel_pcm_intro}, which provides high-level abstractions for monitoring system-wide and core-specific performance metrics, and Intel Perfmon~\cite{perfmon}, a versatile tool for configuring and interacting directly with PMUs. These tools enabled us to easily collect and analyze monitored events. 

  \vspace{-0.1in}
%is a tool designed to reveal the utilization of different modules inside processors. 
%In the Intel 5th Xeon scalable processor family (Emerald Rapids), numerous performance monitor units (PMU\cite{intelpmu}) are integrated to facilitate the comprehensive understanding of processor behaviors from both program and microarchitecture levels. 
%PMUs are specialized hardware components inside a processor that can gather information on metrics such as branch misses, cache misses, CPU stall cycles, memory requests, and so on. Besides the PCM tool, Intel Perfmon\cite{perfmon} provides a set of scripts that enable access to PMU through the perf tool. With PMU's support, users can track and analyze performance bottlenecks in a real-time method.

%\subsection{Methodology}
%Native: no control \\
%1.1 CXL devices (no interleaving) \\
%1.2 CXL (hardware/software) interleaving \\
%1.3 same socket DDR and CXL, cross socket DDR and CXL \\
%1.4 DDR, CXL, and NUMA \\
%1.5 DRAM and CXL interleaving 

\begin{figure*}[t]
\centering
  
  \includegraphics[width=0.92\textwidth]{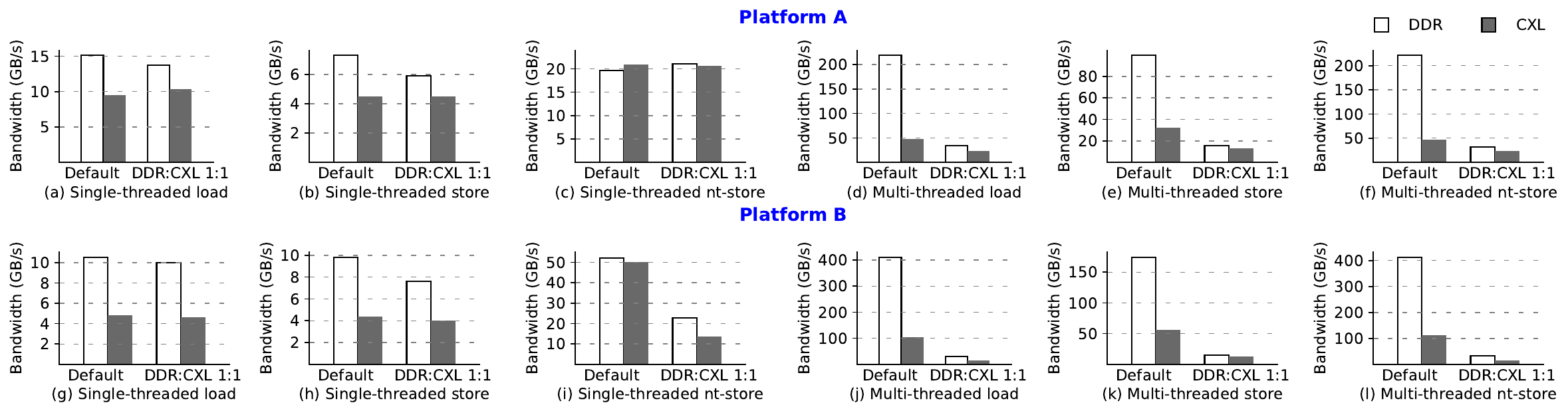}
  \vspace{-0.1in}
  \caption{The comparison of DDR and CXL memory for single-threaded and peak bandwidth on platform A and platform B.}
  \vspace{-0.1in}
  \label{fig:perf-band}
\end{figure*}

% \vspace{-0.2in}

\section{Architectural Implications}
\label{sec:archimpl}
This section examines the architectural impact of CXL memory on the memory hierarchy. Specifically, we investigate how tiered memory's increased latency and performance heterogeneity impact processor efficiency. Modern processors employ various techniques to hide memory latency, such as caching, prefetching, non-blocking caches (i.e., handling multiple concurrent cache misses), and memory-level parallelism (MLP) (i.e., issuing multiple memory requests simultaneously). We take a bottom-up approach to analyze how memory-hiding techniques are affected by memory heterogeneity. Starting with scenarios where all data accesses miss the multi-level caches to assess the impact on MLP, we then examine workloads with strong locality that benefit from CPU caches.

\subsection{Quantifying Performance Heterogeneity}
\label{sec:quant}
While there are studies~\cite{liu2024dissectingcxlmemoryperformance, Sun2023DemystifyingCM} characterizing the performance differences between DDR and CXL memory, we use two commercially available CXL platforms and vary the configurations of DDR and CXL memory to identify the culprits of the performance gap. We investigate whether the performance gap is mainly due to the overhead of the {\tt CXL.mem} protocol, the latency of the PCIe bus, or the internal architecture of CXL memory. Unlike existing work that treats CXL memory as a black-box, standalone device, we aim to understand the architectural differences of CXL memory since its underlying medium is still DDR in current commercial CXL devices. 

Figure~\ref{fig:perf-band} compares DDR and CXL memory regarding single-threaded and peak bandwidth on the platform A and platform B. The benchmarks were single-threaded and multi-threaded {\tt bw-test} running separately on DDR and CXL memory. For each benchmark run, only one type of memory was tested. Platform A has two CPU sockets, each equipped with 8 DDR DIMMs and 2 CXL memory devices, while platform B has a single socket with 12 DDR DIMMs and 4 CXL devices. By default, both platforms have all DDR and CXL memory slots fully populated. All DDR DIMMs were interleaved by hardware while CXL memory devices were interleaved by system software using the weight interleaving~\cite{kernel-interleaving} in Linux with equal weight. Besides the default configurations, we removed all but one DDR and CXL memory, respectively, to test performance without interleaving. 

{\noindent \bf Single-threaded bandwidth}. As expected, CXL memory achieved lower bandwidth than DDR memory for different types of memory instructions. However, the performance gap was not significant for single-threaded bandwidth tests. DDR and CXL memory achieved comparable bandwidth, with DDR outperforming CXL memory by at most $52\%$ and $100\%$ on platform A and platform B, respectively. Notably, a one-to-one (1:1) DDR-to-CXL provisioning ratio helped narrow the gap except for {\tt nt-stores} on the platform B.   

% \textcolor{red}{
We further observe in Figure~\ref{fig:perf-band} (c) that on the platform A, CXL's performance is very close to or slightly surpasses DDR. However, in Figure~\ref{fig:perf-band} (i) on the platform B, the performance gap in the one-to-one (1:1) DDR-to-CXL case is more pronounced compared to the Default configuration. Upon further investigation, we found that a single thread exhibits different capabilities on the platform A and platform B in achieving the maximal memory bandwidth for {\tt nt-stores}. Specifically, on the platform A, a single thread can achieve a maximum memory bandwidth of approximately 20 GB/s. As a result, even though the Default case offers higher hardware parallelism, the limited per-thread capacity causes both the Default and one-to-one cases to achieve similar performance. In contrast, on the platform B, a single thread can achieve much higher {\tt nt-stores} performance -- around 50 GB/s. This allows the Default case to significantly outperform the one-to-one case. The larger performance gap observed in the one-to-one case compared to the Default configuration on the platform B is due to the fact that, although the Default case benefits from greater hardware parallelism, a single thread is still constrained by its own limited capability, preventing it from fully utilizing the available memory bandwidth. This is further confirmed through multi-threaded tests Figure~\ref{fig:perf-band} (f) and (l) (see below): Without the capability limitation from a single thread, DDR outperforms CXL in multi-threaded tests, and the performance gap between the two is smaller in the one-to-one configuration compared to the Default configuration.
% }

{\noindent \bf Multi-threaded bandwidth}. In contrast, multi-threaded bandwidth tests, which provided sufficient thread-level concurrency to leverage memory-level parallelism, resulted in a large discrepancy between DDR and CXL performance. As shown in Figure~\ref{fig:perf-band} (d) - (f) and (j) - (l),  the performance gap between DDR and CXL memory was as large as $4x$ for multi-threaded {\tt load} and {\tt nt-store} instructions on both platforms. Interestingly, the gap significantly narrows in the 1-to-1 DDR-CXL test, where both hardware and software interleaving between DDR and CXL memory were disabled. 
Overall, the performance gap is narrower in the 1-to-1 DDR-CXL test compared to the default configurations, where the DDR-to-CXL DIMM ratios were 8:2 on the platform A and 12:4 on the platform B. On both platforms, there is a strong correlation between multi-threaded bandwidth and DIMM-level parallelism -- the peak bandwidth for all three memory instructions closely approximates the product of the single DIMM performance and the number of DIMMs for both DDR and CXL memory.

\begin{figure}[t]
\centering
  
  \includegraphics[width=0.42\textwidth]{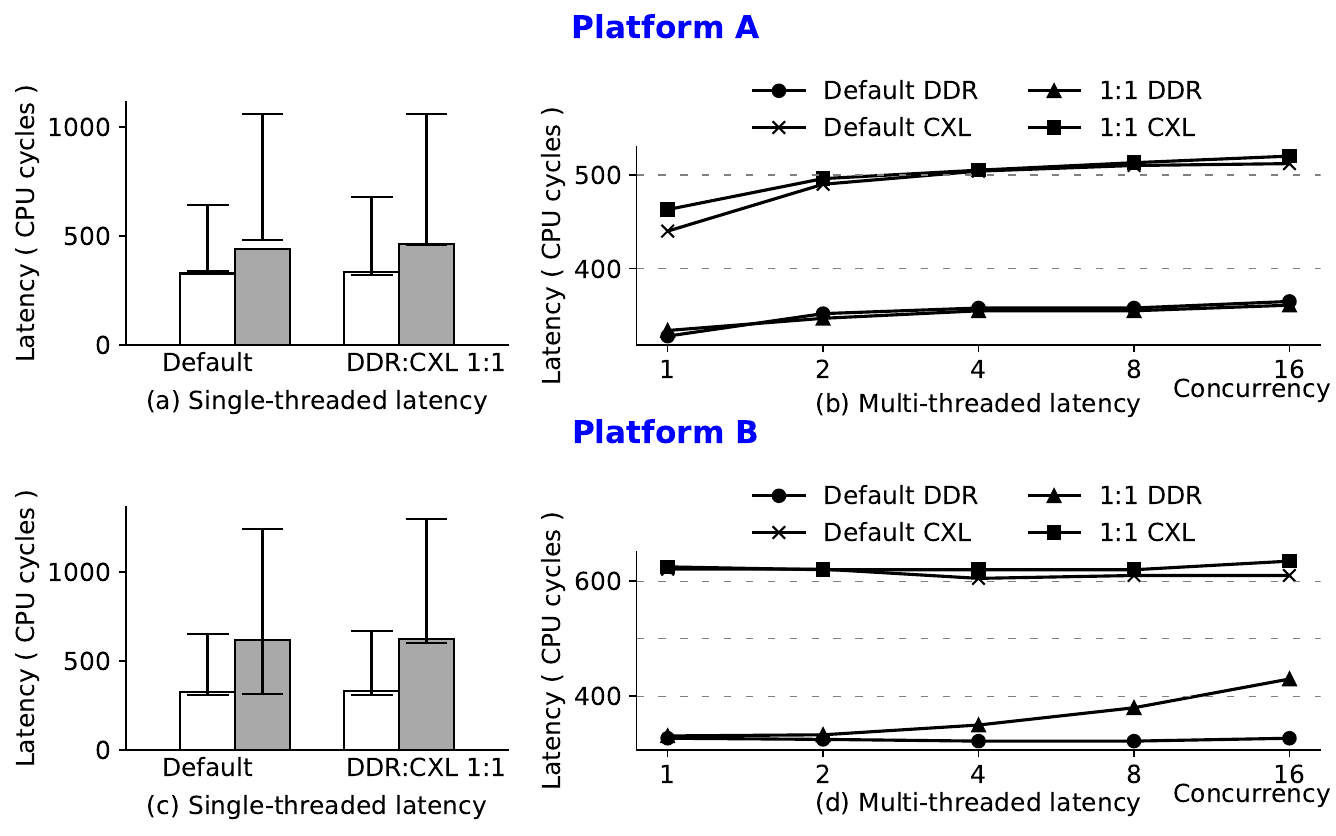}
 \vspace{-0.1in}
  \caption{The comparison of DDR and CXL memory latency.}
  \vspace{-0.2in}
  \label{fig:perf-lat}
\end{figure}
{\noindent \bf Average and tail latency}. The {\tt bw-test} benchmark continuously issues memory instructions to sequentially access memory without involving any computation. Thus, it leverages the instruction-level parallelism in the CPU pipeline, non-blocking caches, and memory-level parallelism to issue a large number of concurrent memory instructions to stress test memory bandwidth. In contrast, the {\tt lat-test} benchmark issues memory instructions to cachelines belonging to a randomly linked list via pointer chasing. The pointer to the next cacheline is stored in the current cacheline being fetched from memory, thereby inhibiting concurrent issues of multiple memory instructions. As a result, the latencies of individual instructions do not overlap and measure the round trip time to complete data access without any queuing delay either within the processor or the memory device. Figure~\ref{fig:perf-lat} shows the average and tail latency of DDR and CXL memory due to {\tt load} instructions. The error bars indicate the median and 99\textsuperscript{th} percentile latency. The WSS of the {\tt lat-test} benchmark was set to 512 MB to ensure most data accesses missed the caches and were served from memory. As shown in Figure~\ref{fig:perf-lat} (a) and (c), while CXL memory exhibits higher average and tail latencies, its performance is not significantly different from that of DDR memory for single-threaded latency test. Additionally, DIMM-level parallelism has no significant impact on latency. In the multi-threaded latency test, the platform A and platform B exhibited different behaviors due to differences in processor design, yet both DDR and CXL demonstrated similar performance trends. Figure~\ref{fig:perf-lat} (b) and (d) show the average latency as concurrency increases. 
The changes in tail latency were within $10\%$ of those observed in the single-threaded test and are thus not shown. Unlike in the bandwidth test, the average latencies of DDR and CXL memory did not change dramatically as concurrency increased.  

{\noindent \bf Analysis.}
The bandwidth and latency tests shed light on the overhead and internal architecture of CXL memory. Compared to DDR memory, CXL memory achieves lower yet comparable performance when the CXL device is not overloaded, i.e., in all latency tests and the single DIMM bandwidth test. The (almost constant) overhead could be attributed to the cost of the {\tt CXL.mem} protocol and the latency of the PCIe bus. Large bandwidth gaps are observed in the default configurations that fully populate the DDR and CXL slots. Although CXL memory devices have larger capacities than DDR memory, its hardware parallelism, e.g., bank-level parallelism, is not greater than that in a single DDR DIMM. 
% \textcolor{red}{Since the two testbeds have fewer CXL slots than DDR slots, the CXL memory operates with fewer memory channels, leading to reduced overall hardware parallelism (i.e., in our experimental testbeds).}
Since the two testbeds have fewer CXL slots than DDR slots, the CXL memory operates with fewer memory channels, leading to reduced overall hardware parallelism (i.e., in our experimental testbeds).

\subsection{Memory Request Handling}
This section investigates how memory performance heterogeneity can affect memory request handling. We focus first on the platform A, as its hardware performance monitoring tools offer deeper insights into architectural events. Findings will also be verified with the platform B. As sufficient concurrency and memory traffic are needed to stress test memory request handling, we used the {\tt bw-test} with multiple threads as the workload. The WSS was set to 32 GB, two orders of magnitude larger than the LLC size (160 MB on the platform A), to bypass all CPU caches.

\begin{figure}[t]
\centering
  
  \includegraphics[width=0.45\textwidth]{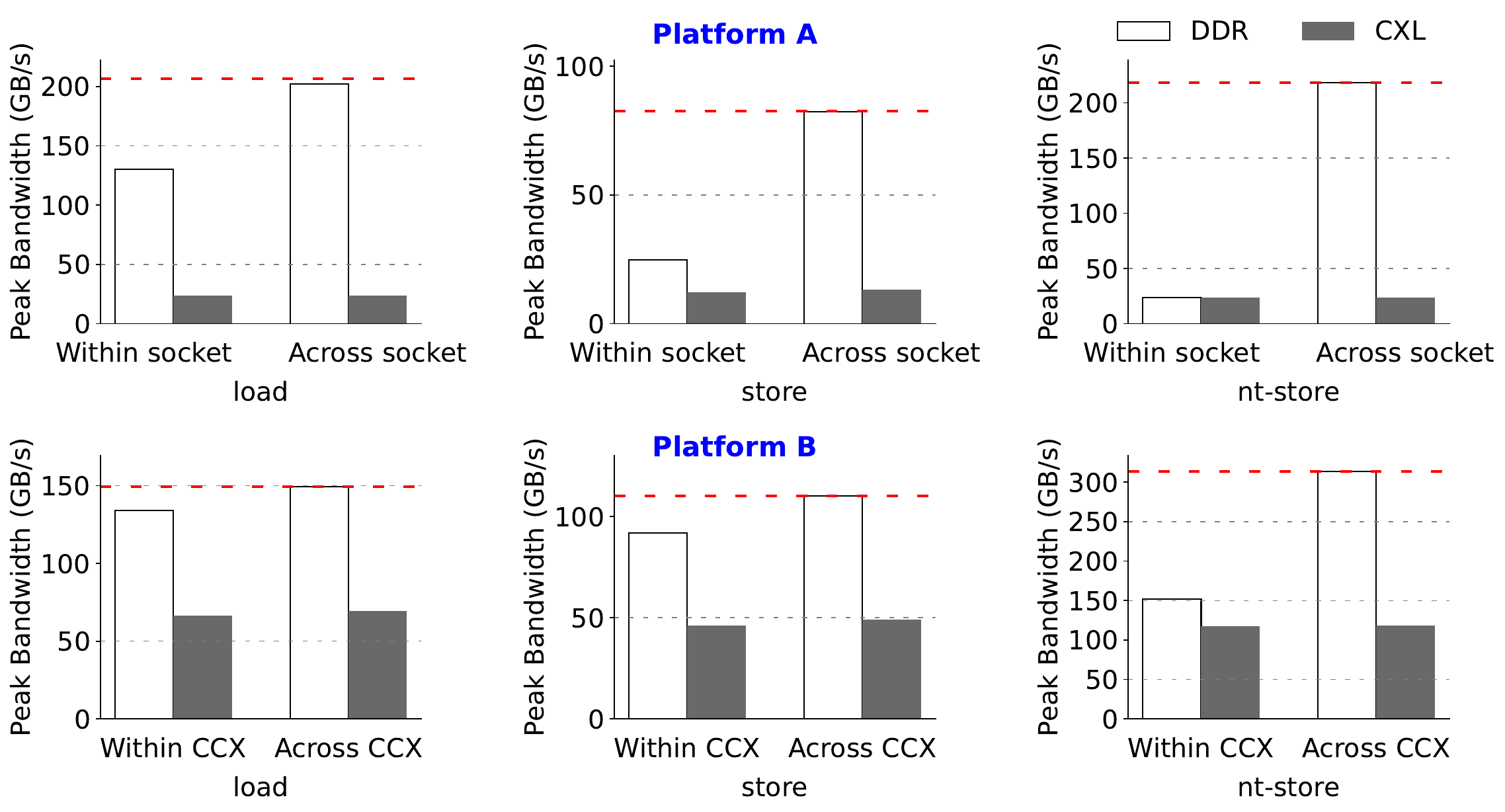}
  % \caption{Significant bandwidth loss due to concurrent handling of DDR and CXL memory requests. \textcolor{red}{The red dotted lines indicate the maximally achievable DDR bandwidth in the absence of CXL memory traffic.}}
  \caption{Significant bandwidth loss due to concurrent handling of DDR and CXL memory requests. The red dotted lines indicate the maximally achievable DDR bandwidth in the absence of CXL memory traffic.}
  \vspace{-0.1in}
  \label{fig:band-tor}
  \vspace{-0.1in}
\end{figure}

\begin{figure}[t]
\centering
  
  \includegraphics[width=0.45\textwidth]{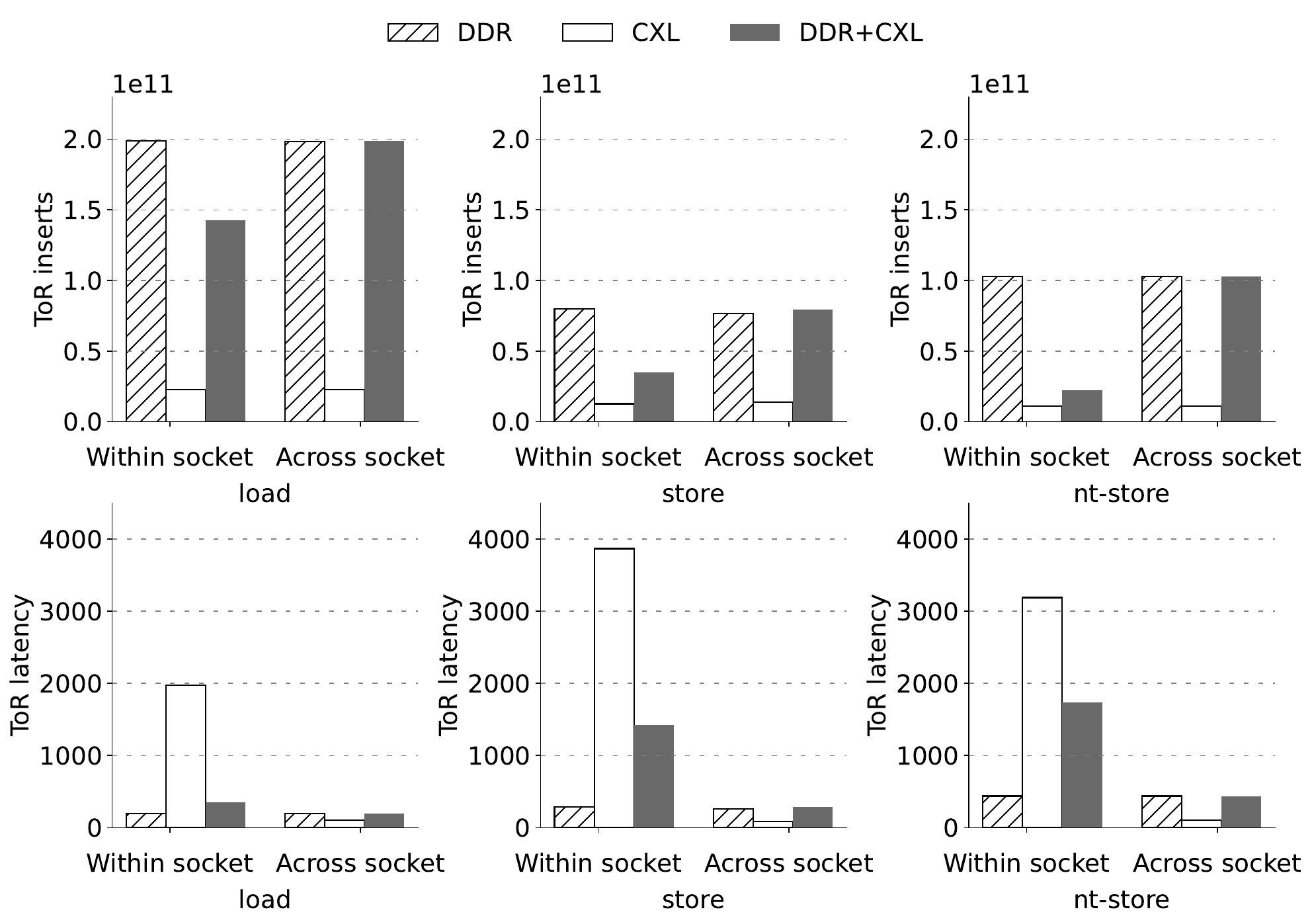}
  \vspace{-0.1in}
  \caption{The insertion rate and average latency of memory requests in ToR.}
  \vspace{-0.2in}
  \label{fig:lat-tor}
\end{figure}

The vendor A's processor employs the Caching and Home Agent (CHA) to manage cache coherence within and across socket (s), and handle memory requests. The CHA is part of the uncore system that includes the LLC, the inter-socket connect, and the memory controller. Any data access that misses the L2 cache or triggers cache coherence operations is managed by the CHA. Incoming memory and coherence requests are first queued in the Ingress Request Queue (IRQ) for initial processing. Requests that require further resolution are admitted to the Table of Requests (ToR) for status tracking and coordination. ToR-managed transactions include LLC hits, coherence operations, LLC misses that result in local and remote (cross-socket) memory accesses, and non-cacheable data accesses such as {\tt non-temporal stores}. For scalability, there are multiple CHAs in a processor, each physically attached to a core in a tile which also includes the core's private caches (L1 and L2), and a slice of the shared LLC. The CHAs are connected to the processor's mesh interconnect and are logically shared among all cores. Data accesses from any core are routed to the corresponding CHA, often located on a different core tile, based on physical address hashing. This ensures an even distribution of the physical address space across CHAs, facilitating efficient communication between cores and memory controllers. ToRs, located within each CHA, function as a unified, logical queue shared among all cores. If not otherwise stated, ToR refers to the logically shared queue across multiple core tiles. 

{\noindent \bf Performance bottleneck}. Figure~\ref{fig:band-tor} shows DDR and CXL memory bandwidth when the processor issues simultaneous memory requests to both memory devices. The benchmarks were two instances of 16-thread {\tt bw-test}, each accessing DDR and CXL memory, respectively. Ideally, system-wide bandwidth should closely approach the combined bandwidth of DDR and CXL since they run on separate buses. However, when the two benchmarks ran within a single socket on the platform A, DDR experienced significant performance degradation, while CXL was only slightly impacted by the co-running. As indicated by the red dotted lines, which represent the maximum achievable DDR bandwidth, DDR's performance loss was as much as $89\%$. Isolating the two benchmarks on separate sockets eliminated DDR bandwidth loss. These observations suggest that a bottleneck within a socket prevents efficient handling of heterogeneous memory requests.  

We further investigate the platform B for a similar issue. While the vendor B's processor only has a single socket, it includes 12 Core Complex (CCX), each of which is a cluster of 7 CPU cores integrated within a die. Like the CHA in Vendor A's processors, CCX is responsible for handling LLC hits, interacting with the memory controller, and ensuring memory requests are properly routed across CCXs and tracked. As shown in Figure~\ref{fig:band-tor}, co-running DDR and CXL workloads within a CCX similarly led to degraded DDR bandwidth, especially for {\tt nt-stores}.  

{\noindent \bf Root cause analysis}. We monitor architectural events using hardware performance counters on the platform A to provide insights into the potential bottleneck. Since ToR is where all memory requests are queued and tracked,  we monitor two hardware events related to memory request handling. {\tt UNC\_CHA\_TOR\_INSERTS.all} measures the total number of ToR insertions across all types of requests. {\tt UNC\_CHA\_TOR\_OCCUPANCY.all} measures the number of active entries in the ToR. Both events are cumulative, and ToR statistics are recorded per CPU cycle. Therefore, the ratio of the two events, i.e., $\frac{\text{\small UNC\_CHA\_TOR\_OCCUPANCY}}{\text{\small UNC\_CHA\_TOR\_INSERTS}}$, reports the average latency in CPU cycles of memory requests that retire from the ToR queue. Figure~\ref{fig:lat-tor} shows the ToR insertion rate and the average request latency in various scenarios. We used the ToR insertion rate during the {\tt bw-test} benchmark running exclusively on DDR memory, achieving the highest DDR bandwidth, as the reference. We compared the reference ToR insertion rate with that observed during DDR and CXL co-running. Since it is not possible to separate ToR insertions for DDR and CXL, the ToR insertion rate for {\tt DDR+CXL} within the same socket included both requests. It is evident that the ToR insertion rate is strongly correlated with the measured memory bandwidth with a Pearson correlation coefficient of 0.998, indicating a strong positive correlation.

Next, we explain how the heterogeneous performance of different memory devices can lead to overall degraded memory performance. Recall that performance heterogeneity is due to the inherent overhead of the CXL protocol and the lack of hardware parallelism in the CXL device. We calculated the average ToR request latency using the ratio of ToR occupancy and inserts. Since any active requests in ToR are already dispatched to memory devices, the ToR request latency is effectively the memory service time. As shown in Figure~\ref{fig:lat-tor}, there is a significant difference in DDR and CXL latency for the bandwidth test measured at the ToR. While DDR latency increased slightly compared to the latency observed in the latency test (Figure~\ref{fig:perf-lat}), CXL latency rose by nearly an order of magnitude for all three memory instructions. In the {\tt DDR+CXL} test, where the ToR handled a mix of DDR and CXL requests, ToR latency rose substantially as the overall memory performance declined. We observed that the memory service time, whether for DDR or CXL memory, rises but remains stable even when the memory device is overloaded, e.g., incurring significant request queuing on the device due to a lack of hardware parallelism. Thus, we can calculate the composition of ToR requests based on DDR and CXL latencies measured offline and the average ToR latency when there are simultaneous DDR and CXL workloads (i.e., case {\tt DDR+CXL}). Our analysis indicated that ToR entries were disproportionately occupied by CXL requests, limiting DDR memory throughput and leading to overall memory bandwidth degradation.

{\noindent {\bf \em How does unfair queue allocation in the ToR arise?}} Our findings reveal a fundamental issue in handling data access to memory with heterogeneous performance. Vendor A's processors equip two queues in the CHA to process data access requests. Besides the ToR, which contains dispatched requests, the IRQ is a staging place where data requests from CPU cores are initially queued. IRQ is responsible for flow control and enforcing priorities between different types of requests. When the IRQ is full, it signals upstream components, such as cores and other CHAs, to slow down the issue of new requests. As DDR and CXL memory have distinctly different service times, especially under heavy load, ToR creates backpressure for requests with high service time in the IRQ. As a result, CXL memory requests gradually accumulate in the IRQ because cores issue DDR and CXL requests to the IRQ at the same rate, even though DDR requests are completed more quickly at the ToR. When the IRQ becomes full, the CHA throttles both DDR and CXL requests from upstream components, perpetuating the disproportionate occupation of the IRQ by CXL requests. This results in an imbalanced request dispatch rate to the ToR, which adversely impacts the performance of the faster DDR memory.

{\noindent \bf Implication \#1}: The performance heterogeneity of tiered memory, particularly the disparity in hardware parallelism, can lead to disproportionate queuing in the CHA on the platform A. This unfair processing of memory requests penalizes access to faster memory, resulting in significant bandwidth degradation. We expect a similar issue in the CCX on the platform B. 

% \textcolor{red}{
Although the high-level principle of Implication \#1 -- i.e., requests to faster memory slow down when queued behind slower ones -- is not unprecedented, we are, to the best of our knowledge, the first to identify and quantify the impact of slower CXL memory on fast DDR performance in emerging tiered memory systems with DDR and CXL memory. Using micro-benchmarks and hardware performance counters, we demonstrate this effect with empirical results. Furthermore, we provide an in-depth architecture-level analysis that uncovers the root cause of this issue on platform A -- i.e., how the limited hardware parallelism of CXL memory and the complex interplay between IRQ and ToR contribute to performance degradation. These insights would offer valuable guidance for future architectural innovations. Finally, we introduce a simple yet practical system-level solution (Section \S\ref{sec:design}) that effectively mitigates this performance bottleneck, ensuring near-peak DDR throughput while still leveraging the benefits of CXL memory.
% }

\subsection{Shared LLC}

\begin{figure*}[t]
\centering
  
  \includegraphics[width=0.8\textwidth]{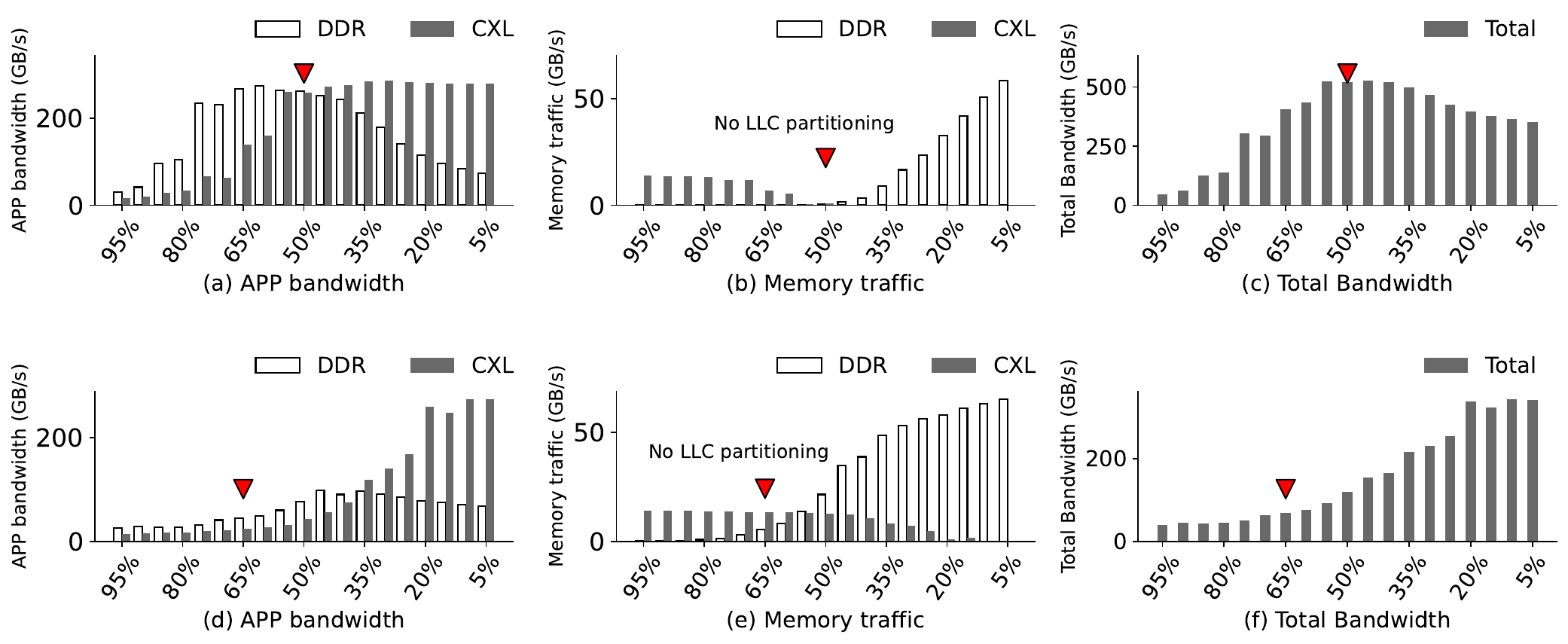}
\vspace{-0.1in}
  % \caption{Memory performance heterogeneity can negatively affect the effectiveness of LLC on the Intel platform. \textcolor{red}{The x-axis represents the amount of LLC allocated to the DDR workload, with the remaining portion allocated to the CXL workload. The red arrow highlights the scenario equivalent to ``free competition'', where no LLC partitioning is enforced between workloads.}}
  \caption{Memory performance heterogeneity can negatively affect the effectiveness of LLC on the platform A. The x-axis represents the amount of LLC allocated to the DDR workload, with the remaining portion allocated to the CXL workload. The red arrow highlights the scenario equivalent to ``free competition'', where no LLC partitioning is enforced between workloads.}
  \vspace{-0.1in}
  \label{fig:llc-partition}
\end{figure*}

This section examines whether the heterogeneous performance of tiered memory impacts the effectiveness of the LLC, which is also managed by the CHA. Data accesses that miss the L2 cache are queued at the ToR for status tracking and performance monitoring. Thus, both LLC hits and misses (outstanding memory requests) are subject to contentions on the ToR queue. To illustrate the severity of the problem, we designed two co-running workloads with strong locality that benefit from LLC caching. Each workload consisted of a {\tt bw-test} using {\tt store} instructions running on DDR memory and CXL memory, respectively. Figure~\ref{fig:llc-partition} shows the performance of the co-running workloads due to heterogeneous memory performance under two configurations. 

We begin with setting each workload's WSS to 60 MB and the combined WSS is below the LLC capacity (i.e., 160 MB in the Vendor A's EMR processor). Both workloads were placed within the same socket and shared the ToR. Since the combined WSS fits comfortably within the LLC, the expected bandwidth for both workloads should closely approach LLC performance, despite the performance gap between DDR and CXL memory. Figure~\ref{fig:llc-partition} (a) - (c) show the performance breakdown of the two workloads with varying LLC allocations, in which the red arrows indicate free competition between the workloads on LLC usage. We used Intel's Cache Allocation Technology (CAT)~\cite{dcat} to partition the cache capacity between the two workloads. As shown in Figure~\ref{fig:llc-partition} (a), under free competition, the LLC allocations between the two workloads were fair, with both DDR and CXL workloads achieving similarly high bandwidth due to the majority of data accesses being LLC hits. Additionally, Figure~\ref{fig:llc-partition} (b) and (c) confirm minimal memory traffic missing the LLC, ensuring that system-wide throughput was maximized under free competition (indicated by red arrows). 

As LLC allocations varied from favoring the DDR workload (left) to promoting the CXL workload (right), we observed distinct performance trends for the DDR and CXL workloads. When the DDR workload was allocated most of the LLC capacity (i.e., $95\%$), its performance unexpectedly degraded significantly, failing to benefit from the LLC even though its entire working set was cached. On the contrary, when the LLC capacity was mostly allocated to the CXL workload, both workloads achieved high performance. The CXL workload achieved a nearly 280 GB/s bandwidth thanks to LLC caching while the DDR workload also scored 80 GB/s, close to the peak DDR bandwidth (w/o caching) for {\tt store} instructions. 

Next, we increased the WSS of each workload to 120 MB -- each can fit entirely in the LLC but the combined WSS exceeds LLC capacity. This setting ensures that each workload can potentially benefit from LLC caching but there are significant memory accesses. As shown in Figure~\ref{fig:llc-partition} (d) - (f), free competition did not result in fair LLC allocation or overall optimal performance. Since CXL memory incurred higher latency, during free LLC competition it took a longer time to serve LLC misses, thereby resulting in a lower LLC fill rate. The DDR workload occupied approximately 65\% of the LLC space but achieved less-than-expected bandwidth, even falling below the DDR memory bandwidth without LLC caching. Increasing the LLC allocation for DDR using CAT further reduced the DDR workload's bandwidth. An examination of ToR hardware events revealed that the contention on ToR entries caused by CXL memory requests significantly slowed down the processing of DDR LLC hits, which are also managed by ToR. In contrast, the maximum system throughput was achieved by allocating the majority of the LLC capacity to the CXL workload to mitigate its high memory latency, as shown in Figure~\ref{fig:llc-partition}(f).

{\noindent \bf Implication \#2}: The disproportionate queuing of memory requests due to memory heterogeneity can also negatively impact LLC performance. The fundamental issue is that LLC hits, memory, and coherence requests are all handled by the caching and home agent (CHA), making them vulnerable to unfair request processing.

\subsection{Private Caches}

\begin{figure}
\centering
  
  \includegraphics[width=0.35\textwidth]{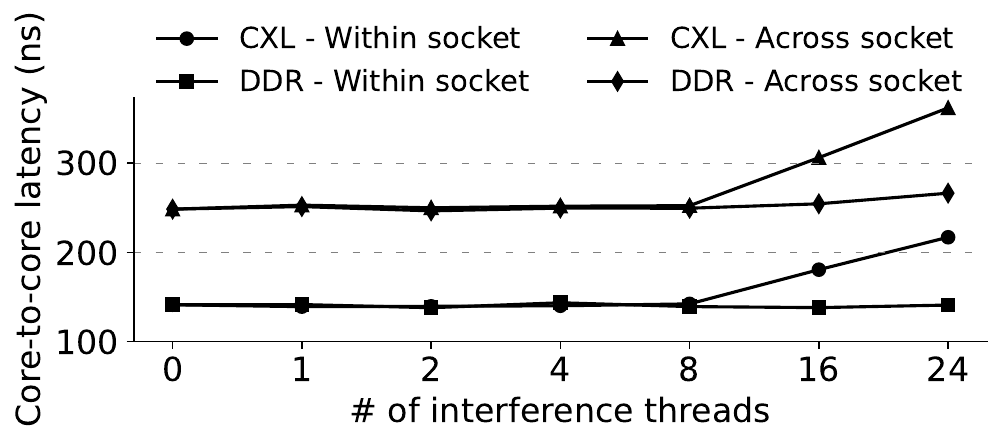}
  \vspace{-0.1in}
  \caption{Data access to CXL memory can affect cross-core synchronization on the platform A.}
  \vspace{-0.2in}
  \label{fig:sharing}
\end{figure}
This section investigates whether unfair request processing in the CHA and similar components can affect the performance of workloads whose WSS fits into the private L1 and L2 caches. Intuitively, data accesses that hit the private caches are handled independently by individual cores, making them immune to interference from other cores. Evaluation with the {\tt bw-test} workload using a small WSS showed no performance issues, regardless of the underlying memory type. However, coherence requests across private caches still need to be handled by the CHA and queued at the ToR. To confirm this potential problem, we created a simple 2-thread benchmark {\tt lat-share} that repeatedly updates a shared cacheline using atomic instruction {\tt CAS}. Background bandwidth-intensive {\tt bw-test} workloads were run along with {\tt lat-share} to cause contentions in the CHA. Figure~\ref{fig:sharing} shows the average latency of each atomic update with interference from either DDR or CXL memory accesses. As CXL interference intensified, synchronization latency rose substantially, while DDR interference had no noticeable impact. Unlike bandwidth-bound tests affected by CXL memory access from within the same socket, {\tt lat-share} is similarly impacted by CXL interference from a different socket.  

{\noindent \bf Implication \#3}: Our evaluation with the platform B revealed that only cross-CCX synchronization is affected by CXL interference, suggesting that coherence requests are handled separately from other memory requests within a CCX. On the platform A, all coherence, memory, and LLC requests are centrally managed by the CHA, thereby vulnerable to CXL interference. 

\section{System Implications and Optimizations}
\label{sec:design}
Section \S\ref{sec:archimpl} reveals that while memory-intensive applications operate on distinct memory tiers on separate buses (i.e., local DDR and remote CXL memory), they contend for shared hardware resources (e.g., the CHA). As suggested in Section \S\ref{sec:archimpl}, uncontrolled DDR and CXL memory interference reduces overall system throughput, increases memory access latency, and ultimately degrades the performance of co-running applications. 

One conventional mitigation strategy is to implement {\em strict isolation} using (hardware/software-level) resource limits or reservations, such as assigning dedicated portions of the ToR to local DDR and remote CXL memory separately. However, static reservations/limits undermine the {\em work-conserving} property, resulting in under-utilization of available hardware resources. This limitation becomes even more pronounced for CXL memory, which, unlike local DDR, is both expandable and scalable. For instance, the capacity and bandwidth of CXL memory can be adjusted on demand by leveraging CXL memory pools~\cite{pond}. Moreover, since CXL memory pools can be shared across multiple hosts, the available memory bandwidth to a single host may fluctuate. Therefore, static ToR allocation for CXL memory restricts its ability to adapt to the inherent flexibility and dynamic scalability of CXL memory. Note that, at the time of writing, there is no hardware support for custom allocation/partitioning ToR for either vendor A or vendor B processors. Instead, a dynamic memory request control scheme is desired, capable of differentiating and managing requests from tiered, heterogeneous memory devices to maximize overall system performance. 
%Given that tiered memory systems typically store performance-critical data (e.g., hot data) in the fast tier, such a scheme should prioritize serving requests to local DDR, effectively leveraging its low latency and high bandwidth. At the same time, remote CXL memory requests should be handled on a best-effort basis, utilizing the additional capacity and bandwidth. In essence, when managing a mix of memory requests, the dynamic scheme must prioritize local DDR requests within the constraints of limited hardware resources (e.g., ToR) while using any remaining resources to handle CXL memory requests. Recognizing the dynamic scheme requires hardware and system software support.

\subsection{Memory Request Control Knobs}
\label{sec:knobs}
In this section, we introduce existing key hardware and software features and mechanisms that can help control memory requests and their limitations in achieving a dynamic memory control scheme. 

\noindent{\bf Memory bandwidth allocation.} Modern processors from Intel and AMD offer advanced hardware features for memory resource management, e.g., Intel's Resource Director Technology (RDT)~\cite{closerlook} and AMD's Platform Quality of Service (QoS)~\cite{amdqos}. These technologies enable monitoring and control over shared resources, e.g., LLC and memory bandwidth. Intel's Memory Bandwidth Allocation (MBA) technology, as part of RDT, provides per-core control over memory bandwidth. This is achieved by regulating the memory request rates from a core's private L2 cache to the shared LLC. 

The MBA technology leverages the concept of Classes of Service (CLOS), which defines specific throttling parameters for shared memory bandwidth. Multiple cores can be assigned to the same CLOS, thereby subjecting them to the bandwidth limitations configured for that class. Further, system software, such as operating systems (OS), is responsible for managing the association between running threads and their respective CLOS. Specifically, when a software thread is scheduled to run on a core (or logical processor), the OS updates the \texttt{IA32\_PQR\_ASSOC} Model Specific Register (MSR) to reflect the CLOS assigned to that running thread. This update enables the hardware to enforce the appropriate memory bandwidth limit for the thread based on its configured settings.

Intuitively, in tiered memory systems, MBA can be leveraged to prioritize local DDR by throttling memory requests of applications that access CXL memory -- by regulating CXL memory requests at an earlier stage (i.e., from L2 to LLC), the number of CXL memory requests subsequently inserted into the ToR can be controlled.

\noindent{\bf \em Limitations.} While MBA's throttling operates as a {\em static} percentage of available memory bandwidth and can ensure strict performance isolation by throttling memory bandwidth for ``noisy'' neighbors (e.g., memory-intensive VMs in the public cloud), it lacks the ability to dynamically adapt to changing workload demands. This limitation is compounded by the absence of system software support to detect whether local DDR requests are under-served and whether throttling of CXL requests is necessary. Determining the level of throttling for CXL requests while avoiding performance interference between local DDR and remote CXL memory and ensuring work-conserving behavior remains a challenge.

%In addition, we observed, consistent with previous studies~\cite{hostcongestioncc, closerlook}, that the existing MBA mechanism to control memory requests remains {\em coarse-grained}. It offers a limited and discrete set of throttle levels (e.g., 10 throttle levels). MBA control is also not immediately responsive. While the throttling parameter can be dynamically and rapidly adjusted by writing to the MBA MSR (within tens of microseconds), it takes a relatively longer delay before the bandwidth throttling becomes fully effective (e.g., up to seconds).

\noindent{\bf Memory interleaving.} CXL memory bandwidth expansion through heterogeneous memory interleaving is highlighted as a key value proposition for CXL-attached memory devices~\cite{micron-perf}. The latest Intel and AMD processors support {\em hardware memory interleaving} within the same type of memory devices, DDR or CXL memory\footnote{Currently, Intel EMR processors do not support hardware interleaving for CXL memory, whereas AMD Genoa 9634 includes this capability.}, but do not support interleaving across DDR and CXL memory. Instead, as introduced in Section~\S\ref{sec:archimpl}, {\em software-based memory interleaving} in heterogeneous memory systems could potentially enhance overall memory throughput by distributing consecutive memory addresses across local DRAM and CXL memory at the page level. An optimal interleaving ratio is critical to prevent memory tier overloads and maximize bandwidth, as imbalances can cause request backlogs at the ToR and degrade overall system performance. The optimal interleaving ratio depends on factors like the relative bandwidth of DDR and CXL memory, the number of memory channels, and the workload's memory access patterns. 

\noindent{\bf \em Limitations.} Determining an optimal interleaving ratio is difficult with dynamic or unpredictable CXL memory and workloads, where the capability of CXL memory and access patterns of workloads can vary significantly over time. 
Further, we observed that the optimal interleaving ratio changes when the bandwidth ratio between local DDR and CXL memory is altered due to factors such as increased DDR speed or a different number of CXL channels. While a dynamic adaptation or workload-aware interleaving ratio configuration mechanism can mitigate this issue (to a certain degree), it involves complex system software components that control page allocation policies and interleaving algorithms. Moreover, these changes do not affect pages already allocated or need to cause interruptions to running workloads for page re-allocation, leading to downtime or performance disruptions. Last, while memory interleaving could offer increased capacity and bandwidth, it introduces higher latency than local DDR. For many latency-sensitive workloads, placing data fully in local DDR might be more suitable.

\begin{figure}[t]
\centering
  \includegraphics[width=0.35\textwidth]{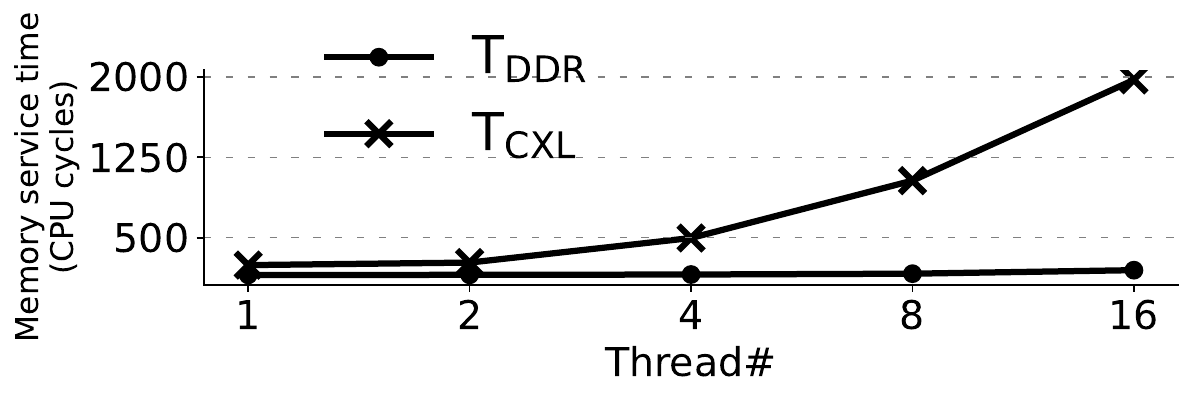}
  \vspace{-0.1in}
  \caption{Average memory service time for DDR and CXL memory with increasing number of threads.}
  \vspace{-0.1in}
  \label{fig:two-sep}
\end{figure}
\subsection{Dynamic Memory Request Control} 

% \textcolor{red}{
% To address the limitations and challenges discussed in Section~\S\ref{sec:knobs}, we introduce a 
% \underline{D}ynamic \underline{M}emory \underline{R}equest \underline{C}ontrol mechanism, named {\em \sys}.  \sys operates on the premise that applications or threads can access any available memory resources, including local DDR, CXL memory, or both.

To address the limitations and challenges discussed in Section~\S\ref{sec:knobs}, we introduce a 
\underline{M}emory \underline{I}nterference \underline{K}nob \underline{U}nit mechanism, named {\em \sys}.  \sys operates on the premise that applications or threads can access any available memory resources, including local DDR, CXL memory, or both.
% }
%It ensures that local DDR requests are prioritized and served promptly, while CXL memory requests are handled on a best-effort basis with low priority if contentions on the shared ToR are observed. 

%\sys dynamically throttles incoming memory requests based on the observed {\em memory service time} tracked by the ToR.  \sys continuously monitors how long memory requests are being tracked in the ToR. Under normal conditions, requests are handled quickly by the LLC or memory devices, resulting in low average service time. In contrast, an increase in the average service time indicates that the system is experiencing a heavy memory load and that the memory devices may be struggling to keep up. In response, \sys dynamically throttles the rate of incoming memory requests to alleviate the pressure, ensuring that the system remains efficient and preventing potential throughput loss. We note that both local DDR and CXL memory could theoretically become bottlenecks due to limited concurrency or capability (e.g., bandwidth). Note that, in practice, local DDR offers significantly higher concurrency and bandwidth compared to CXL memory. Our observations in Section \S\ref{sec:archimpl} also reveal that performance issues arise from CXL memory rather than local DDR. 
%Therefore, our focus is on managing CXL memory traffic to sustain high system performance.

\noindent{\bf 
% Detecting inefficient memory request handling.} \textcolor{red}{As discussed in Section~\ref{sec:archimpl}, disproportionate memory request queuing in Intel's CHA and AMD's CCX can lead to significant throughput loss, reduced LLC effectiveness, and increased cross-core synchronization overhead. Although memory system overload can lead to similar performance degradation, unfair or inefficient request handling for heterogeneous memory exhibits unique characteristics in low-level hardware events. Memory overloading in homogeneous memory systems leads to a high rejection rate at the ToR queue, while the average request latency at ToR remains relatively stable or increases consistently across various memory requests. In contrast, unfair memory request handling results from significant latency disparities in heterogeneous memory and the backlog of slow memory requests.
Detecting inefficient memory request handling.} As discussed in Section~\ref{sec:archimpl}, disproportionate memory request queuing in platform A's CHA and platform B's CCX can lead to significant throughput loss, reduced LLC effectiveness, and increased cross-core synchronization overhead. Although memory system overload can lead to similar performance degradation, unfair or inefficient request handling for heterogeneous memory exhibits unique characteristics in low-level hardware events. Memory overloading in homogeneous memory systems leads to a high rejection rate at the ToR queue, while the average request latency at ToR remains relatively stable or increases consistently across various memory requests. In contrast, unfair memory request handling results from significant latency disparities in heterogeneous memory and the backlog of slow memory requests.
As illustrated in Figure~\ref{fig:two-sep}, DDR and CXL memory not only exhibit different baseline latencies but also demonstrate distinct latency behaviors in response to increasing request rates.

% \textcolor{red}{
Based on this insight, \sys determines whether CXL memory requests need to be throttled to mitigate unfair request processing by monitoring the latencies of DDR and CXL memory requests at the ToR. If the latency of CXL memory requests exceeds a predefined threshold and continues to grow exponentially, which indicates backpressure from the CXL memory device, \sys starts to throttle CXL memory accesses. 
More concretely, \sys uses hardware performance counters (e.g., Intel's PMU~\cite{intelpmu}) to track two ToR events -- the number of queued requests in the ToR (i.e., \texttt{ToR.Occupancy}) and the number of requests inserted into the ToR (i.e., \texttt{ToR.Inserts}).  According to Little's Law: 
% }

% \textcolor{red}{
\begin{equation}
\label{eq:lat}
T_{\text{avg}} = \frac{\text{ToR.Occupancy}}{\text{ToR.Inserts}} = \alpha\% \cdot T_{\text{ddr}} + (1 - \alpha\%) \cdot T_{\text{cxl}}
\end{equation}
% }

% \textcolor{red}{
The average memory service time of all memory requests at ToR ($T_{avg}$) can be measured by $\texttt{ToR.Occupancy} / \texttt{ToR.Inserts}$. As shown in Equation (1), $T_{avg}$ further comprises a mix of requests from DDR, accounting for $\alpha$\%, and from CXL memory, accounting for (1-$\alpha$\%).  \sys calculates $\alpha$ by tracking read and write instructions to DDR and CXL memory via {\tt uncore} hardware events. According to micro-benchmark {\tt bw-test}, {\tt load}, {\tt store}, and {\tt nt-store} instructions overload CXL memory at different levels of concurrency. Ordinary {\tt store} instructions are read-modify-write operations that involve an equal number of reads and writes while {\tt nt-store} instructions only involve writes. Through extensive benchmarking, we discovered that the latency of CXL write requests (i.e., {\tt nt-stores}) is roughly twice that of read requests (i.e., {\tt loads}) at the same concurrency level. Additionally, the latency of {\tt stores} can be approximated as the average latency of read and write requests. Our profiling also identified a throttling threshold for read latency, beyond which CXL memory throughput declines, and the queuing delay on CXL memory devices increases exponentially~\footnote{The write latency threshold is approximately twice that of the read latency.}. Specifically, we use the measured $T_{avg}$ and $\alpha$, in addition to $T_{DDR}$ which is considered constant measured offline~\footnote{In all experiments, DDR memory never caused a backlog in the ToR.}, to calculate $T_{cxl}$ based on Equation (1). Further, we compared $T_{cxl}$ with a target CXL latency, calibrated using the read-write composition of CXL memory requests and their respective latency thresholds measured offline. If $T_{cxl}$ exceeds this threshold, a backlog of CXL requests builds up in the ToR, leading to inefficient memory request handling. Consequently, CXL request throttling is initiated.

{\noindent \bf Identifying problematic applications.} Once unfair/inefficient memory request handling is detected, \sys identifies applications that are responsible for CXL memory accesses and throttles them to mitigate the memory request backlog. The challenges are twofold. First, hardware performance counters that monitor DDR and CXL requests are system-wide {\tt uncore} events, making them unassociated with any specific application or PID. Second, application access patterns can vary over time, requiring real-time, dynamic tracking of memory accesses. To address these challenges, \sys devises a multi-step procedure to associate CXL memory accesses to PIDs: \ding{182} \sys uses the memory bandwidth monitor (MBM) feature of the Intel RDT technology to locate cores with high memory traffic that most likely contribute to the CXL request backlog. We empirically define memory traffic of 10 MB/s or higher as ``high''. \ding{183} \sys uses core IDs to traverse their run queues to obtain a list of threads (PIDs) that run on these busy cores. \ding{184} \sys uses processor-event-based-sampling (PEBS) to sample memory instructions and their virtual addresses associated with each PID. \ding{185} Finally, \sys translates the virtual addresses of these memory accesses to physical addresses based on PID's page tables. Since DDR and CXL memory reside in separate regions of the physical address space, the sampled physical addresses are sufficient to identify PIDs that access CXL memory. To dynamically detect CXL access, \sys updates the CXL-accessing PID list every second.

% \noindent{\bf Effective CXL request throttling.} \textcolor{red}{\sys employs a hierarchical throttling mechanism that integrates coarse-grained application isolation with fine-grained memory request control. To enable isolation, \sys classifies CPU cores into two groups: {\em unrestricted} and {\em restricted}. DDR requests and CXL requests during a period with no CXL backlog can run freely on unrestricted cores. Restricted cores are where CXL-accessing threads are isolated to control their memory access precisely. Furthermore, restricted cores provide sufficient hardware-level parallelism for applications to sustain peak CXL memory bandwidth without causing a request backlog. Since {\tt load}, {\tt store}, and {\tt nt-store} instructions saturate CXL memory bandwidth at different concurrency levels, the restricted cores are further categorized into three levels, each corresponding to the maximum backlog-free concurrency level for these instructions. We empirically determined the three levels (level-1 to 3) to include 8, 4, and 1 cores, respectively, for {\tt load}, {\tt store}, and {\tt nt-store}. }

\noindent{\bf Effective CXL request throttling.} \sys employs a hierarchical throttling mechanism that integrates coarse-grained application isolation with fine-grained memory request control. To enable isolation, \sys classifies CPU cores into two groups: {\em unrestricted} and {\em restricted}. DDR requests and CXL requests during a period with no CXL backlog can run freely on unrestricted cores. Restricted cores are where CXL-accessing threads are isolated to control their memory access precisely. Furthermore, restricted cores provide sufficient hardware-level parallelism for applications to sustain peak CXL memory bandwidth without causing a request backlog. Since {\tt load}, {\tt store}, and {\tt nt-store} instructions saturate CXL memory bandwidth at different concurrency levels, the restricted cores are further categorized into three levels, each corresponding to the maximum backlog-free concurrency level for these instructions. We empirically determined the three levels (level-1 to 3) to include 8, 4, and 1 cores, respectively, for {\tt load}, {\tt store}, and {\tt nt-store}.

% \textcolor{red}{
Upon the detection of CXL request backlog, i.e., the measured $T_{cxl}$ exceeds its threshold, \sys moves all threads accessing CXL memory to level-3, the most restrictive level with only one core. This is to ensure that the request backlog is promptly resolved. If $T_{cxl}$ still fails to meet its target, \sys further throttles the memory access rate at level 3, either directly using Intel's MBA bandwidth control or indirectly by adjusting the CPU quota in the Linux control group (cgroup)~\cite{cgroup}. Restricting an application's CPU time can effectively regulate its memory access rate. Otherwise, \sys gradually promotes threads to less restrictive levels until $T_{cxl}$ approaches its threshold, maximizing concurrency for CXL-accessing threads while ensuring minimal impact on DDR accesses. Note that the throttling process involves multiple rounds of adjustment before $T_{cxl}$ stabilizes. 
% }

% {\noindent \bf Overhead analysis.} \textcolor{red}{
% \sys requires per-second sampling windows to provide reliable detection of CXL memory access. Our measurements across various workloads indicate that stabilizing $T_{cxl}$ and achieving the maximum allowable CXL bandwidth takes approximately 6–7 seconds. Thus, \sys is most effective for workloads that alternate between DDR and CXL memory accesses every tens of seconds or longer. \textcolor{red}{This limitation is largely due to a lack of hardware support (e.g., performance counters) for efficiently detecting and differentiating various types of memory accesses and accurately associating them with their corresponding threads. }}

{\noindent \bf Overhead analysis.} 
\sys requires per-second sampling windows to provide reliable detection of CXL memory access. Our measurements across various workloads indicate that stabilizing $T_{cxl}$ and achieving the maximum allowable CXL bandwidth takes approximately 6–7 seconds. Thus, \sys is most effective for workloads that alternate between DDR and CXL memory accesses every tens of seconds or longer. This limitation is largely due to a lack of hardware support (e.g., performance counters) for efficiently detecting and differentiating various types of memory accesses and accurately associating them with their corresponding threads.

\vspace{-0.1in} 
\section{Case Studies}
We evaluated the effectiveness of \sys using micro-benchmarks and real-world applications as case studies. For micro-benchmarks, we used the same {\tt bw-test} benchmarks as described in Section~\S\ref{sec:archimpl}. Additionally, we evaluated \sys with three representative memory-intensive real-world applications: a large language model (LLM) serving service running on CPUs~\cite{cpullm}, a big data processing application~\cite{ApacheSpark}, and an in-memory key-value caching~\cite{dashmap}.
%\noindent{\bf Baselines for comparison.} 
We compared {\em four} cases. 1) {\em DataRacing}: Workloads ran simultaneously on local DDR and CXL memory without any throttling. 2) {\em \sys}: Dynamic throttling of workloads accessing CXL memory using CPU quotas. 3) {\em \sys-MBA}: Dynamic throttling of workloads accessing CXL memory using Intel's MBA feature; we replaced CPU quota based memory request control with Intel's MBA, while still leveraging \sys's profiling and 2-phase adjustment mechanism. 4) {\em Opt}: Workloads ran separately on DDR or CXL memory, eliminating performance interference. We ran these experiments with the same configurations as stated in Section~\S\ref{sec:profiling}.

% \noindent{\bf Micro-benchmarks.}
% Same as Section~\S\ref{sec:archimpl}, we ran two {\tt bw-test} workloads with 16 threads to stress test tiered memory systems: one on local DDR and the other on remote CXL memory, to simulate a data racing scenario. \textcolor{red}{Further, to simulate a more dynamic scenario, these threads alternate DDR and CXL access every 100 seconds. As shown in Figure~\ref{fig:eval-micro-switch}, both \sys and \sys-MBA enable the \texttt{store} and \texttt{nt-store} test cases on DDR to achieve performance very close to the optimal case, meanwhile allowing all test cases (i.e., \texttt{load}, \texttt{store}, and \texttt{nt-store}) to significantly outperform the data racing scenario. This improvement occurs because \sys's  a two-level isolation/throttling approach allows for prompt mitigation of CXL request backlog while enabling CXL-bound workloads to use maximum allowable concurrency without causing a backlog. We also observe that both CPU quotas and MBA are effective for workloads that alternate between DDR and CXL access at a slower rate (e.g., 100 seconds) compared to \sys's stabilizing speed (e.g., 6-7 seconds), resulting in similar performance.}

\begin{figure}[t]
\centering
  \includegraphics[width=0.45\textwidth]{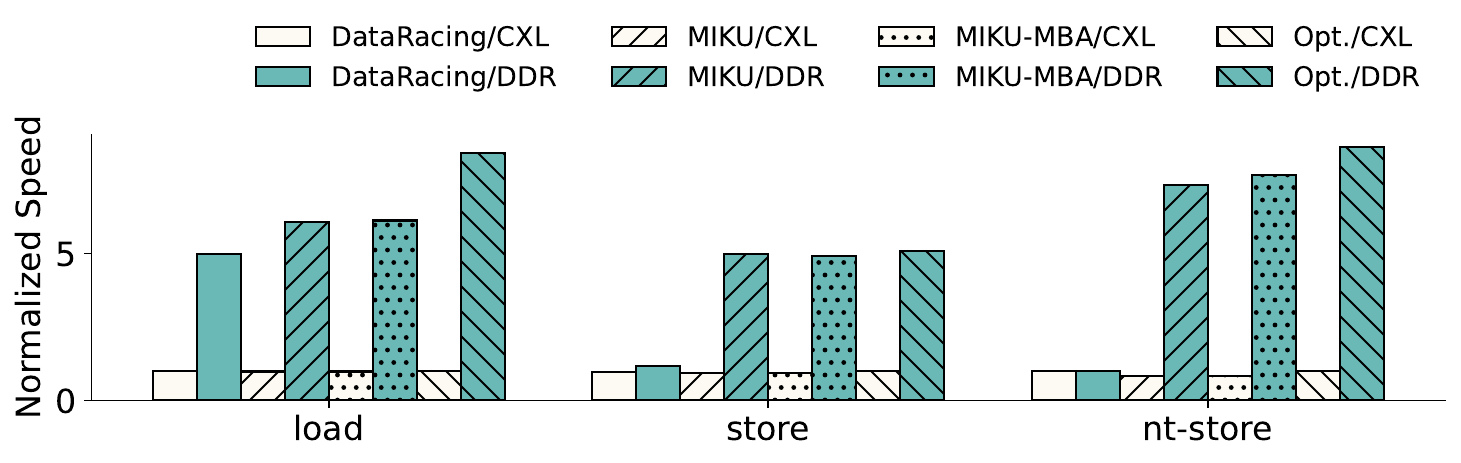}
  \vspace{-0.1in}
  % \caption{\textcolor{red}{Performance comparisons with microbenchmarks, which alternate DDR and CXL access every 100 seconds.}}
  \caption{Performance comparisons with microbenchmarks, which alternate DDR and CXL access every 100 seconds.}
  \vspace{-0.1in}
  \label{fig:eval-micro-switch}
\end{figure}

\noindent{\bf Micro-benchmarks.}
Same as Section~\S\ref{sec:archimpl}, we ran two {\tt bw-test} workloads with 16 threads to stress test tiered memory systems: one on local DDR and the other on remote CXL memory, to simulate a data racing scenario. Further, to simulate a more dynamic scenario, these threads alternate DDR and CXL access every 100 seconds. As shown in Figure~\ref{fig:eval-micro-switch}, both \sys and \sys-MBA enable the \texttt{store} and \texttt{nt-store} test cases on DDR to achieve performance very close to the optimal case, meanwhile allowing all test cases (i.e., \texttt{load}, \texttt{store}, and \texttt{nt-store}) to significantly outperform the data racing scenario. This improvement occurs because \sys's  a two-level isolation/throttling approach allows for prompt mitigation of CXL request backlog while enabling CXL-bound workloads to use maximum allowable concurrency without causing a backlog. We also observe that both CPU quotas and MBA are effective for workloads that alternate between DDR and CXL access at a slower rate (e.g., 100 seconds) compared to \sys's stabilizing speed (e.g., 6-7 seconds), resulting in similar performance.

%prioritize DDR requests while handling CXL requests on a best-effort basis. 
%As the {\tt bw-test} workloads do not exhibit much dynamics, we do not observe much difference between \sys and sys-MBA. However, as we will see shortly, with more workload dynamics, \sys outperforms \sys-MBA due to more timely, fine-grained memory request control.
%Show the effectiveness of our dynamic memory request control scheme
%"Run Alone", "Co-locate w/o control", "PCM-based control", and "RDT-based control"
%1) For memory bandwidth intensive workloads, our approach can prioritize DDR requests while serving CXL requests on a best-effort maner, thereby improving overall memory performance.
%2) For workloads that under-utilizes memory bandwidth, our approach introduces negligible overhead.

% \begin{figure}[t]
% \centering
%   \includegraphics[width=0.45\textwidth]{acmart-primary/samples/figures/microbench.pdf}
%   \vspace{-0.1in}
%   \caption{Performance comparisons with microbenchmarks.}
%   \vspace{-0.2in}
%   \label{fig:eval-micro}
% \end{figure}

\begin{figure}[t]
\centering
  \includegraphics[width=0.45\textwidth]{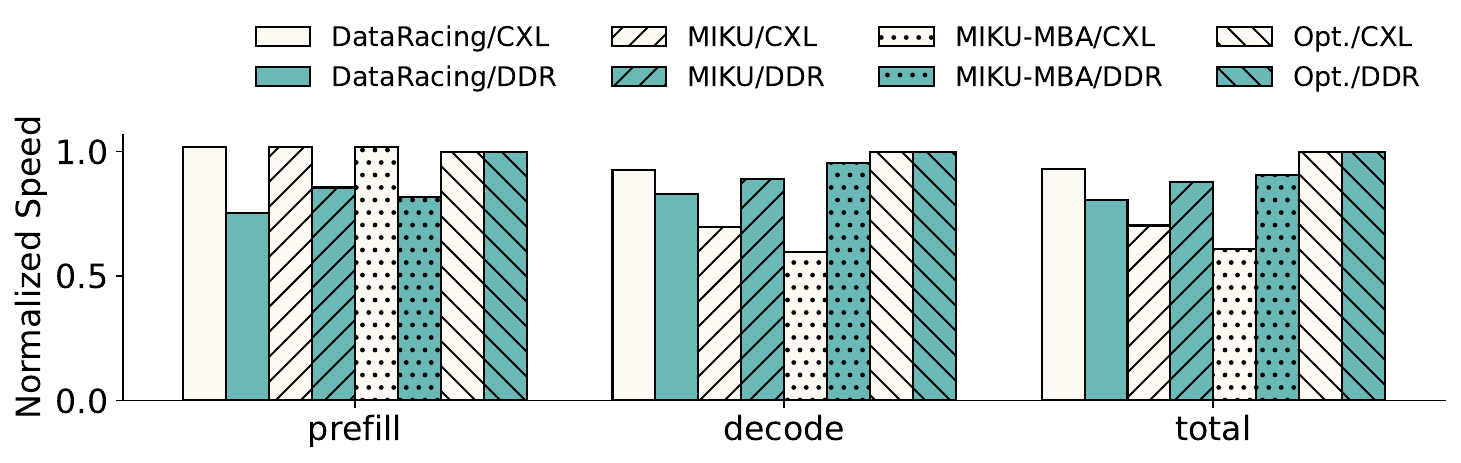}
  \vspace{-0.1in}
  \caption{Performance comparisons with LLM inferences. The results of experiments on DDR and CXL are normalized to their respective optimal performance result.}
  \vspace{-0.2in}
  \label{fig:llm}
\end{figure}

\begin{figure*}[t]
\centering
  \includegraphics[width=0.98\textwidth]{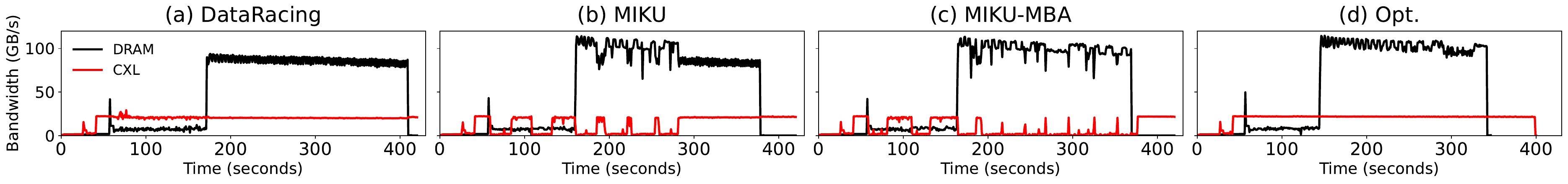}
  \vspace{-0.1in}
  \caption{LLM serving: Memory bandwidth changes over the different phases of LLM serving.}
  \vspace{-0.15in}
  \label{fig:llama}
\end{figure*}

\begin{figure*}[t]
\centering
  \includegraphics[width=0.98\textwidth]{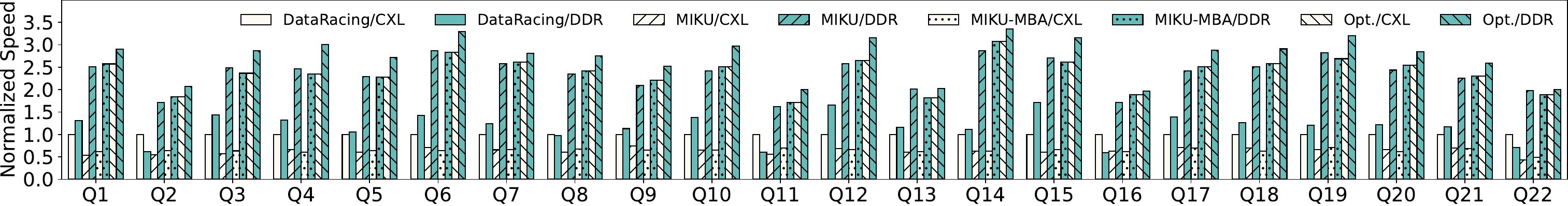}
  \vspace{-0.1in}
  \caption{Spark reads and analyzes database tables to serve 22 TPC-H business-oriented queries.}
  \vspace{-0.2in}
  \label{fig:spark-tpch}
\end{figure*}

\noindent{\bf LLM serving.} We next evaluated the end-to-end performance of Large Language Model (LLM) inference workloads~\cite{llamacpp,cpullm}. A typical LLM inference process comprises two stages: 1) {\em Prefill}: During this {\em compute-intensive} stage, the LLM processes input tokens and generates the first token. 2) {\em Decode}: In this stage, the LLM auto-regressively generates tokens one at a time until reaching the stop conditions. The {\em decode} stage is highly {\em memory-bandwidth-intensive for reads} and often causes memory bandwidth degradation for co-located LLM inference instances. To simulate data racing, we ran two LLM inference instances concurrently: a LLaMA 3.1 70B q4 quantized model on local DDR memory and a LLaMA 3.1 8B fp16 model on a single CXL memory module. We configured the input length, output length, and batch size to 128, 512, and 4, respectively.

% As illustrated in Figure \ref{fig:llm}, under data racing, the performance of the bandwidth-intensive {\em decode} stage on DDR memory drops to 70\% of its optimal performance. However, by limiting the CPU quota, thereby throttling memory requests for LLM inference on CXL memory, \textcolor{red}{\sys achieves 89\% of the optimal {\em decode} performance.}

As illustrated in Figure \ref{fig:llm}, under data racing, the performance of the bandwidth-intensive {\em decode} stage on DDR memory drops to 70\% of its optimal performance. However, by limiting the CPU quota, thereby throttling memory requests for LLM inference on CXL memory, \sys achieves 89\% of the optimal {\em decode} performance.

We further conducted an in-depth analysis of the dynamic memory bandwidth behavior over time during the experiments. As shown in Figure~\ref{fig:llama} (d), when two instances run independently, the peak DRAM bandwidth during the {\em decode} stage can reach up to 120 GB/s. In contrast, the CXL instance reaches the maximum bandwidth capacity of a single CXL memory module, resulting in device overloading. Figure~\ref{fig:llama} (a) illustrates the memory bandwidth behavior when the two instances run concurrently (without throttling). During the {\em decode} stage of the DRAM instance, its memory bandwidth is limited to 75 GB/s due to contention for hardware resources caused by the overloaded CXL device.
%which degrades DRAM performance. 
Figure~\ref{fig:llama} (b) highlights the effectiveness of \sys in mitigating this issue. When the {\em decode} stage of the DRAM instance begins, \sys detects the increased memory latency and dynamically limits the CPU quota of the CXL instance. Once the DRAM instance completes its decode stage, \sys restores the CPU quota of the CXL instance to its default settings, ensuring resource efficiency. Similar behaviors are also observed under \sys's MBA mechanism in Figure~\ref{fig:llama} (c).

\noindent{\bf Big data processing.} 
We further conducted our experiments using Apache Spark~\cite{ApacheSpark}, a high-performance engine for big data processing. We generated a 100GB dataset based on TPC-H~\cite{TPCH}, an industry-standard benchmark for SQL queries. Spark processes data through pipelines of read-only in-memory Resilient Distributed Datasets (RDDs), which involve intensive memory allocation, data shuffling, and impose substantial memory stress.
We ran two Spark processes to simulate data racing when multiplexing system resources. Each process was allocated 20GB of memory for Spark executors and 40GB for Spark drivers. One process operated entirely on a local DDR memory node, while the other utilized a single local CXL memory module. 
% Figure \ref{fig:spark-tpch} shows that 1) Under data racing, Spark's performance on DDR memory drops to 30\% of its optimal speed (Q2), while the performance on CXL memory decreases to 32\% (Q14). 2) By limiting the CPU quota, which in turn throttles memory requests, for Spark on CXL memory, \sys achieves at least 81\% of the optimal performance on DDR, while the performance on CXL memory is reduced to 45\%–75\% compared to the data racing scenario. \textcolor{red}{3) We observed that \sys-MBA achieves similar performance as \sys CPU quotas}.
Figure \ref{fig:spark-tpch} shows that 1) Under data racing, Spark's performance on DDR memory drops to 30\% of its optimal speed (Q2), while the performance on CXL memory decreases to 32\% (Q14). 2) By limiting the CPU quota, which in turn throttles memory requests, for Spark on CXL memory, \sys achieves at least 81\% of the optimal performance on DDR, while the performance on CXL memory is reduced to 45\%–75\% compared to the data racing scenario. 3) We observed that \sys-MBA achieves similar performance as \sys CPU quotas. 
%to restore Spark's performance due to its slow response, which cannot adapt to Spark's dynamically fluctuating memory speed.

\begin{figure}[t]
\centering
  \includegraphics[width=0.45\textwidth]{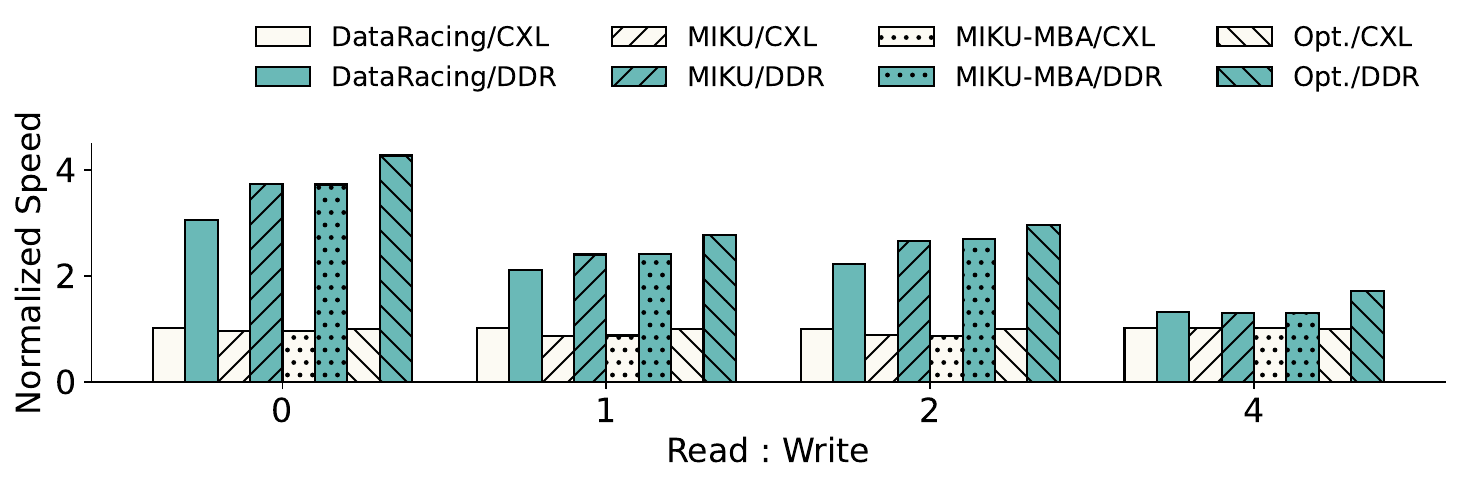}
  \vspace{-0.1in}
  \caption{Key-value caching via a concurrent hash table.}
  \vspace{-0.25in}
  \label{fig:hashtable}
\end{figure}
\noindent{\bf In-memory key-value caching.} 
% Finally, we evaluated an industry-standard key-value store dataset using a concurrent hash table. Specifically, we selected dashmap~\cite{dashmap}, a high-performance, lock-free concurrent map implemented in the Rust programming language. The dataset was generated using YCSB~\cite{ycsb}, a widely used benchmark for cloud-serving systems. To simulate data racing, we ran two concurrent hashmaps: one on local DDR memory and the other on CXL memory, with a working set size (WSS) of approximately 7 GB. The operations consisted of two types: {\tt get} for reads and {\tt insert} for writes, with the read-to-write ratio varied between 0 (pure insert) and 4. \textcolor{red}{As shown in Figure~\ref{fig:hashtable}, both \sys and \sys-MBA achieve better performance than the {\em Dataracing} scenario, while the performance gap narrows as there are more read operations (e.g., with a read-to-write ratio from 0 to 4). This result is because, like the {\tt lat-test} (\S\ref{sec:archimpl}), concurrent hashmaps involve random memory accesses that do not fully saturate DDR memory bandwidth, leading to less frequent activation of memory control mechanisms. Additionally, a higher ratio of \texttt{inserts}, which involve more complex read-modify-write operations than \texttt{reads}, generally results in a greater memory workload, allowing \sys to demonstrate its effectiveness more.}

Finally, we evaluated an industry-standard key-value store dataset using a concurrent hash table. Specifically, we selected dashmap~\cite{dashmap}, a high-performance, lock-free concurrent map implemented in the Rust programming language. The dataset was generated using YCSB~\cite{ycsb}, a widely used benchmark for cloud-serving systems. To simulate data racing, we ran two concurrent hashmaps: one on local DDR memory and the other on CXL memory, with a working set size (WSS) of approximately 7 GB. The operations consisted of two types: {\tt get} for reads and {\tt insert} for writes, with the read-to-write ratio varied between 0 (pure insert) and 4. As shown in Figure~\ref{fig:hashtable}, both \sys and \sys-MBA achieve better performance than the {\em Dataracing} scenario, while the performance gap narrows as there are more read operations (e.g., with a read-to-write ratio from 0 to 4). This result is because, like the {\tt lat-test} (\S\ref{sec:archimpl}), concurrent hashmaps involve random memory accesses that do not fully saturate DDR memory bandwidth, leading to less frequent activation of memory control mechanisms. Additionally, a higher ratio of \texttt{inserts}, which involve more complex read-modify-write operations than \texttt{reads}, generally results in a greater memory workload, allowing \sys to demonstrate its effectiveness more.

\vspace{-0.1in}
\section{Discussions}
% \noindent{\bf Tiered memory provisioning}. \textcolor{red}{This paper demonstrates that the limited hardware-level parallelism in CXL memory is a primary cause for its significantly higher latency, which can potentially compromise overall system efficiency. While this paper does not explicitly focus on the mechanisms of workload partitioning across the tiered memory, it highlights that partitioning a workload's memory to match the hardware parallelism of CXL memory with that of local DDR memory should be a key consideration, potentially even more critical than memory capacity, in tiered memory provisioning. However, challenges emerge in efficiently provisioning CXL memory from shared memory pools to multiple hosts with varying degrees of parallelism in their local memory and eventually to workloads with dynamic memory access patterns.}

\noindent{\bf Tiered memory provisioning}. This paper demonstrates that the limited hardware-level parallelism in CXL memory is a primary cause for its significantly higher latency, which can potentially compromise overall system efficiency. While this paper does not explicitly focus on the mechanisms of workload partitioning across the tiered memory, it highlights that partitioning a workload's memory to match the hardware parallelism of CXL memory with that of local DDR memory should be a key consideration, potentially even more critical than memory capacity, in tiered memory provisioning. However, challenges emerge in efficiently provisioning CXL memory from shared memory pools to multiple hosts with varying degrees of parallelism in their local memory and eventually to workloads with dynamic memory access patterns.

% \noindent{\bf Data access patterns in tiered memory}. \textcolor{red}{Although workloads' data can span across DDR and CXL memory, the access pattern is affected by tiered memory provisioning and management. We consider three scenarios and discuss whether \sys's dynamic CXL access detection and throttling will be effective. First, application data can be statically allocated and pinned in CXL memory. For these workloads, \sys can avoid CXL request backlog as long as their execution time last more than tens of seconds, allowing \sys to detect a backlog and stabilize the access adjustment. Second, application data can be initially allocated to DDR and is later demoted to CXL memory due to OS-level page management (e.g., transparent page placement (TPP)~\cite{Maruf23-tpp}) based on access recency and frequency. As the data residing in CXL memory is deemed cold, threads that frequently alternate DDR and CXL accesses in this scenario typically show poor locality and involve significant indexing or pointer chasing operations. While \sys may not detect CXL access timely, as discussed in Section~\ref{sec:quant}, these workloads are unlikely to cause CXL backlogs. Third, alternating DDR and CXL accesses can also be due to memory interleaving, where frequent CXL accesses that cause a backlog can be addressed by configuring an appropriate interleaving ratio between DDR and CXL.}

\noindent{\bf Data access patterns in tiered memory}. Although workloads' data can span across DDR and CXL memory, the access pattern is affected by tiered memory provisioning and management. We consider three scenarios and discuss whether \sys's dynamic CXL access detection and throttling will be effective. First, application data can be statically allocated and pinned in CXL memory. For these workloads, \sys can avoid CXL request backlog as long as their execution time last more than tens of seconds, allowing \sys to detect a backlog and stabilize the access adjustment. Second, application data can be initially allocated to DDR and is later demoted to CXL memory due to OS-level page management (e.g., transparent page placement (TPP)~\cite{Maruf23-tpp}) based on access recency and frequency. As the data residing in CXL memory is deemed cold, threads that frequently alternate DDR and CXL accesses in this scenario typically show poor locality and involve significant indexing or pointer chasing operations. While \sys may not detect CXL access timely, as discussed in Section~\ref{sec:quant}, these workloads are unlikely to cause CXL backlogs. Third, alternating DDR and CXL accesses can also be due to memory interleaving, where frequent CXL accesses that cause a backlog can be addressed by configuring an appropriate interleaving ratio between DDR and CXL.

\noindent{\bf Architectural innovations}. The fundamental issue in processing heterogeneous memory requests within the processor is the lack of memory type-specific flow control to manage and prioritize requests for different memory tiers. Architectural innovations, such as token-based differential request scheduling, warrant investigation.

\noindent{\bf Application insights}. Our profiling revealed that modern processors can handle memory requests with higher latency without sacrificing overall throughput, provided the underlying memory devices are not overloaded, which will result in exponentially increased latency and a severe backlog of requests within the processor. This finding sheds light on the types of applications that are most suitable for running with CXL memory, despite its high latency. Applications involving primarily random memory accesses or inter-access dependencies, such as hash tables and log-structured merge (LSM) trees, are likely to benefit from CXL-based memory expansion without significant performance degradation.

\vspace{-0.11in}

\section{Related Work}

\noindent \textbf{CXL characterization.} Recent studies~\cite{liu2024dissectingcxlmemoryperformance, cxl_study} have investigated the performance characteristics of CXL devices, using both FPGA-based CXL prototypes and commercial ASIC-based CXL products. A key finding from ~\cite{cxl_study} reveals that true CXL memory achieves lower latency and higher bandwidth efficiency compared to emulated CXL memory, showing promise as memory bandwidth expanders. \cite{liu2024dissectingcxlmemoryperformance} provides a more detailed analysis of CXL-induced slowdowns, linking them to increased stall cycles in the CPU backend and reduced cache prefetch efficiency. In this paper, we develop a more comprehensive understanding of the architectural interactions between CXL memory, the host processor, and the memory hierarchy.
%The analysis highlights that the ratio of LLC-stalled cycles, accounting for memory-level parallelism, can be accurately used to predict DRAM-related slowdowns. This insight is used to develop a performance prediction model, which can be leveraged to optimize memory management strategies like NUMA interleaving and memory tiering.

\noindent \textbf{Tiered memory.} In addition to CXL memory~\cite{cxl}, new memory and storage devices, such as high bandwidth memory\cite{hbm}, persistent memory\cite{optane}, and byte-addressable NVMe SSDs\cite{pcieovercxl} are emerging. State-of-the-art tiered memory systems, such as TPP~\cite{Maruf_23asplos_tpp}, Memtis~\cite{memtis}, Nimble~\cite{Yan_19asplos_nimble}, and AutoTiering~\cite{Jonghyeon_21fast}, use memory tiering to manage data across different memory levels. Unlike traditional DRAM-disk hierarchies, these systems treat CXL memory as an extension of DRAM, avoiding redundancy but requiring data movement for optimal access in an exclusive manner.
%where data only resides in one memory tier. 
%As all tiers are byte-addressable and their performance converges, the efficiency of exclusive tiering, given the cost of data movement, remains uncertain. 
In contrast, a more recent work, Nomad~\cite{xiang2024nomad}, proposes a {non-exclusive} strategy, keeping copies of pages on both fast and slow memory tiers to reduce memory thrashing. This paper provides architectural implications benefiting the design and optimization of tiered memory systems.

\noindent {\bf Distributed shared memory (DSM)} systems simplify parallel computing by allowing multiple computers to share memory across a network as a unified space~\cite{190521, 375407, 6121280, lite, farm}. As traditional DSMs are built on non-coherent, high-latency interconnects like Ethernet or InfiniBand, synchronization and massive data transfers for maintaining coherence make DSMs costly.
%suffer from high communication latency 
%due to data transfers between hosts. Techniques like one-sided RDMA over fast interconnects (e.g., InfiniBand) reduce transfer latency by bypassing remote CPUs~\cite{190521, lite}. 
%However,especially for small messages~\cite{farm}. 
Recent works~\cite{partialfailure, 298535, 10597882, mahar2024telepathic} leverage CXL-based shared memory to reduce coherence overhead, using either software-based coherence~\cite{298535} or existing cache coherence protocols from SMP systems. This paper highlights that hardware coherence can be adversely affected by tiered memory systems, calling for further architectural and system research.

% \noindent {\bf Dynamic resource management} \textcolor{red}{techniques, such as limitation and reservation~\cite{cgroup}, have been widely adopted to ensure fair resource sharing~\cite{267124, lu2015vfair, 266915} and performance isolation~\cite{196290, 190517}, to meet desired Quality-of-Service (QoS)~\cite{munikar2022prism}, and to achieve efficient, fine-grained power and energy management~\cite{powercontainers}. Complementary to these existing approaches, this paper contributed new throttling mechanims in the context of emerging tiered memory systems.}

\noindent {\bf Dynamic resource management} techniques, such as limitation and reservation~\cite{cgroup}, have been widely adopted to ensure fair resource sharing~\cite{267124, lu2015vfair, 266915} and performance isolation~\cite{196290, 190517}, to meet desired Quality-of-Service (QoS)~\cite{munikar2022prism}, and to achieve efficient, fine-grained power and energy management~\cite{powercontainers}. Complementary to these existing approaches, this paper contributed new throttling mechanims in the context of emerging tiered memory systems.

\vspace{-0.1in}

\section{Conclusions}
This paper investigates the architectural interactions of CXL memory in the conventional memory hierarchy. Profiling results suggest that 1) the high and unpredictable CXL memory latency is mainly due to the limited hardware parallelism on CXL devices; 2) the performance heterogeneity of tiered memory can cause unfair request processing in the processor, resulting in severely degraded system throughput, ineffective caching, and inefficient synchronization. To address these issues, we develop \sys, a dynamic memory request control mechanism that dynamically adjusts memory request rates based on service time estimates, allowing for prompt mitigation of CXL request backlog while enabling CXL-bound workloads to use maximum allowable concurrency without causing a backlog.
%prioritizing DDR requests while serving CXL requests on a best-effort basis.

\bibliographystyle{ACM-Reference-Format}
\bibliography{reference, reference-cxl}

%\appendix

\end{document}